\documentclass[fleqn,usenatbib]{mnras}

\usepackage{newtxtext,newtxmath}
\usepackage{orcidlink}

\usepackage[T1]{fontenc}

\DeclareRobustCommand{\VAN}[3]{#2}
\let\VANthebibliography\thebibliography
\def\thebibliography{\DeclareRobustCommand{\VAN}[3]{##3}\VANthebibliography}

\usepackage{graphicx}	
\usepackage{amsmath}	
\usepackage{natbib}
\usepackage{booktabs}
\usepackage{pdflscape}

\newcommand{\tess}[1]{\emph{TESS}#1}

\newcommand{\teff}[1]{$T_{\text{eff}}$#1}
\newcommand{\prot}[1]{$P_{\text{rot}}$#1}
\newcommand{\logrhk}[1]{$\log{R'_{\text{HK}}}$#1}
\newcommand{\logg}[1]{$\log{g}$#1}
\newcommand{\vsini}[1]{$v\sin{i}$#1}

\newcommand{\mps}[1]{m s$^{-1}$#1}

\newcommand{\nrv}[1]{145#1}
\newcommand{\degree}{$^{\circ}$}
\defcitealias{Demory_2020}{D20}
\defcitealias{Stefansson_2020}{S20}

\title[A chain of sub-Neptunes around TOI-1266]{Masses, Revised Radii, and a Third Planet Candidate in the ``Inverted'' Planetary System Around TOI-1266}

\author[R. Cloutier et al.]
{Ryan Cloutier,$^{\orcidlink{0000-0001-5383-9393},1,2}$\thanks{E-mail: ryan.cloutier@mcmaster.ca}
Michael Greklek-McKeon,${\orcidlink{0000-0002-0371-1647},^3}$ 
Serena Wurmser,$^{\orcidlink{0000-0003-4791-5331},2}$  
Collin Cherubim,$^{\orcidlink{0000-0002-8466-5469},4,2}$  
\newauthor
Erik Gillis,$^{\orcidlink{0009-0000-4040-6628},1}$  
Andrew Vanderburg,$^{\orcidlink{0000-0001-7246-5438},5}$ 
Sam Hadden,$^{6}$  
Charles Cadieux,$^{\orcidlink{0000-0001-9291-5555},7}$  
\'Etienne Artigau,$^{\orcidlink{0000-0003-3506-5667},7,8}$  
\newauthor
Shreyas Vissapragada,$^{\orcidlink{0000-0003-2527-1475},2}$ 
Annelies Mortier,$^{\orcidlink{0000-0001-7254-4363},9}$ 
Mercedes L\'opez-Morales,$^{\orcidlink{0000-0003-3204-8183},2}$  
David W. Latham,$^{\orcidlink{0000-0001-9911-7388},2}$  
\newauthor
Heather Knutson,$^{\orcidlink{0000-0002-5375-4725},3}$ 
Rapha\"elle D. Haywood,$^{\orcidlink{0000-0001-9140-3574},10}$  
Enric Pall\'e,$^{11}$  
Ren\'e Doyon,$^{\orcidlink{0000-0001-5485-4675},7,8}$  
Neil Cook,$^{\orcidlink{0000-0003-4166-4121},7}$  
\newauthor
Gloria Andreuzzi,$^{12}$
Massimo Cecconi,$^{12}$
Rosario Cosentino,$^{12}$
Adriano Ghedina,$^{\orcidlink{0000-0003-4702-5152},12}$
Avet Harutyunyan,$^{12}$
\newauthor
Matteo Pinamonti,$^{\orcidlink{0000-0002-4445-1845},13}$
Manu Stalport,$^{14}$
Mario Damasso,$^{\orcidlink{0000-0001-9984-4278},13}$ 
Federica Rescigno,$^{\orcidlink{0000-0002-0594-7805},10}$ 
\newauthor
Thomas G. Wilson,$^{\orcidlink{0000-0001-8749-1962},15}$
Lars A. Buchhave,$^{\orcidlink{0000-0003-1605-5666},16}$  
David Charbonneau,$^{\orcidlink{0000-0002-9003-484X},2}$
Andrew Collier Cameron,$^{\orcidlink{0000-0002-8863-7828},15,17}$  
\newauthor
Xavier Dumusque,$^{18}$  
Christophe Lovis,$^{18}$  
Michel Mayor,$^{18}$  
Emilio Molinari,$^{\orcidlink{0000-0002-1742-7735},19}$  
Francesco Pepe,$^{18}$  
\newauthor
Giampaolo Piotto,$^{20}$  
Ken Rice,$^{\orcidlink{0000-0002-6379-9185},21,22}$  
Dimitar Sasselov,$^{\orcidlink{0000-0001-7014-1771},2}$  
Damien S\'egransan,$^{18}$  
Alessandro Sozzetti,$^{\orcidlink{0000-0002-7504-365X},13}$  
\newauthor
St\'ephane Udry,$^{18}$  
Chris A. Watson$^{\orcidlink{0000-0002-9718-3266},23}$  
\\
\\
Affiliations are listed at the end of the paper
}

\pubyear{2022}

\begin{document}
\label{firstpage}
\pagerange{\pageref{firstpage}--\pageref{lastpage}}
\maketitle

\begin{abstract}
  Is the population of close-in planets orbiting M dwarfs sculpted by thermally driven escape or is it a direct outcome of the planet formation process? A number of recent empirical results strongly suggest the latter. However, the unique architecture of the TOI-1266 system presents a challenge to models of planet formation and atmospheric escape given its seemingly ``inverted'' architecture of a large sub-Neptune ($P_b=10.9$ days, $R_{p,b}=2.62\pm 0.11\, \mathrm{R}_{\oplus}$) orbiting interior to that of the system's smaller planet ($P_c=18.8$ days, $R_{p,c}=2.13\pm 0.12\, \mathrm{R}_{\oplus}$). Here we present revised planetary radii based on new TESS and diffuser-assisted ground-based transit observations, and characterize both planetary masses using a set of \nrv{} radial velocity measurements from HARPS-N ($M_{p,b}=4.23\pm 0.69\, \mathrm{M}_{\oplus}, M_{p,c}=2.88\pm 0.80\, \mathrm{M}_{\oplus}$). Our analysis also reveals a third planet candidate ($P_d=32.3$ days, $M_{p,d}\sin{i} = 4.59^{+0.96}_{-0.94}\, \mathrm{M}_{\oplus}$), which if real, would form a chain of near 5:3 period ratios, although the system is likely not in a mean motion resonance. Our results indicate that TOI-1266 b and c are among the lowest density sub-Neptunes around M dwarfs and likely exhibit distinct bulk compositions of a gas-enveloped terrestrial ($X_{\mathrm{env},b}=5.5\pm 0.7$\%) and a water-rich world (WMF$_c=59\pm 14$\%), which is supported by hydrodynamic escape models. If distinct bulk compositions are confirmed through atmospheric characterization, the system's unique architecture would represent an interesting test case of inside-out sub-Neptune formation at pebble traps.
\end{abstract}



\section{Introduction}
The radius valley is a stark feature in the exoplanet population between $\sim 1.6-1.9\, \mathrm{R}_{\oplus}$ that separates terrestrial super-Earths from larger sub-Neptunes, which are enveloped in a low density volatile material such as H/He or water. The radius valley exists around FGK \citep[e.g.][]{Fulton_2017,Fulton_2018,VanEylen_2018,Berger_2020b,Hardegree_2020,Petigura_2022} and M dwarfs alike \citep[e.g.][]{Cloutier_Menou,Hsu_2020,VanEylen_2021}, albeit with distinct dependencies on planet radius and instellation that are suggestive of different radius valley emergence mechanisms around different spectral types. 

Proposed radius valley emergence models include thermally driven mass loss (TDML) of planetary atmospheres, which may be driven by XUV photons from the host star \citep[e.g. photoevaporation;][]{Owen_2013,Jin_2014,Lopez_2014} or from heating by the planetary core after formation \citep[e.g. core-powered mass loss;][]{Ginzburg_2018,Gupta_2019}. Alternatively, the radius valley may emerge directly from the planet formation process via either late-stage super-Earth formation in a gas-depleted environment \citep{Lopez_2018}, by limited gas accretion onto low mass cores \citep{Lee_2015,Lee_2021}, or by water-rich formation followed by inward migration \citep{Raymond_2018,Venturini_2020,Burn_2021}.

A number of lines of recent empirical evidence are suggesting that the M dwarf radius valley likely emerges directly from the planet formation process without the need to invoke atmospheric escape. Planet occurrence rates show evidence that the slope of the M dwarf radius valley in radius-instellation space is inconsistent with TDML models and instead favours a gas-depleted formation scenario \citep{Cloutier_Menou}. The XUV spectrum of the multi-planet host K2-3 cannot explain the low density compositions of its inner planets using TDML models \citep{DiamondLowe_2022}. A sub-population of water-rich planets around M dwarfs may have been uncovered following a uniform reanalysis of archival radial velocity time series \citep{Luque_2022}, although this claim has been contended \citep{Rogers_2023}. The low bulk density of the intensively studied $1.5\, \mathrm{R}_{\oplus}$ planet Kepler-138 d was shown to be inconsistent with TDML and requires a substantial volatile mass fraction suggestive of water-rich formation beyond the snowline \citep{Piaulet_2023}. Lastly, the bulk compositions of seven keystone planets (i.e. within the M dwarf radius valley) require substantial volatile mass fractions and thus disfavour a TDML explanation \citep{Cherubim_2023}. This collection of results paints the emerging picture that unlike around FGK stars, close-in planets around M dwarfs are unlikely to be sculpted by TDML, and instead, we may be witnessing the emergence of a new channel of small planet formation.

TOI-1266 is a validated planetary system from NASA's Transiting Exoplanet Survey Satellite mission that features two small planets purportedly sitting on opposing sides of the M dwarf radius valley \citep[][hereafter \citetalias{Demory_2020} and \citetalias{Stefansson_2020}, respectively]{Demory_2020,Stefansson_2020}. The system was reported to harbour an inner sub-Neptune ($P_b=10.9$ days, $R_{p,b}=2.37\, \mathrm{R}_{\oplus}$) alongside an outer super-Earth \citepalias[$P_c=18.8$ days, $R_{p,c}=1.56\, \mathrm{R}_{\oplus}$;][]{Demory_2020}. Multi-planet system architectures that contain a likely terrestrial planet (i.e. $\lesssim 1.6\, \mathrm{R}_{\oplus}$) whose orbit is wider than the system's sub-Neptune are somewhat rare \citep[$\sim 35$\%;][]{Weiss_2018}. Most planet pairs that span the radius valley exhibit the super-Earth on an orbit that is interior to that of the sub-Neptune. Seemingly ``inverted'' architectures like TOI-1266 are particularly interesting because they can be challenging to reconcile with models of atmospheric escape and population synthesis models based on core accretion theory \citep{Owen_2017,Gupta_2019,Burn_2021}. TOI-1266 is therefore an important testbed for planet formation models, particularly the emergence of the M dwarf radius valley. This motivates our study to measure the planet masses and bulk compositions, and test the consistency of our findings with various radius valley emergence models. 

In this paper we present the results of our campaign to measure the masses of TOI-1266 b and c using precise radial velocity measurements taken with the HARPS-N spectrograph. In Section~\ref{sect:obs} we describe our observations. In Section~\ref{sect:star} we summarize our knowledge of the host star. In Section~\ref{sect:analysis} we present our data analysis. In Section~\ref{sect:disc} we discuss our results and their implications for sub-Neptune formation around M dwarfs before concluding with a summary of our key findings in Section~\ref{sect:summary}.

\begin{figure}
    \centering
    \includegraphics[width=0.85\hsize]{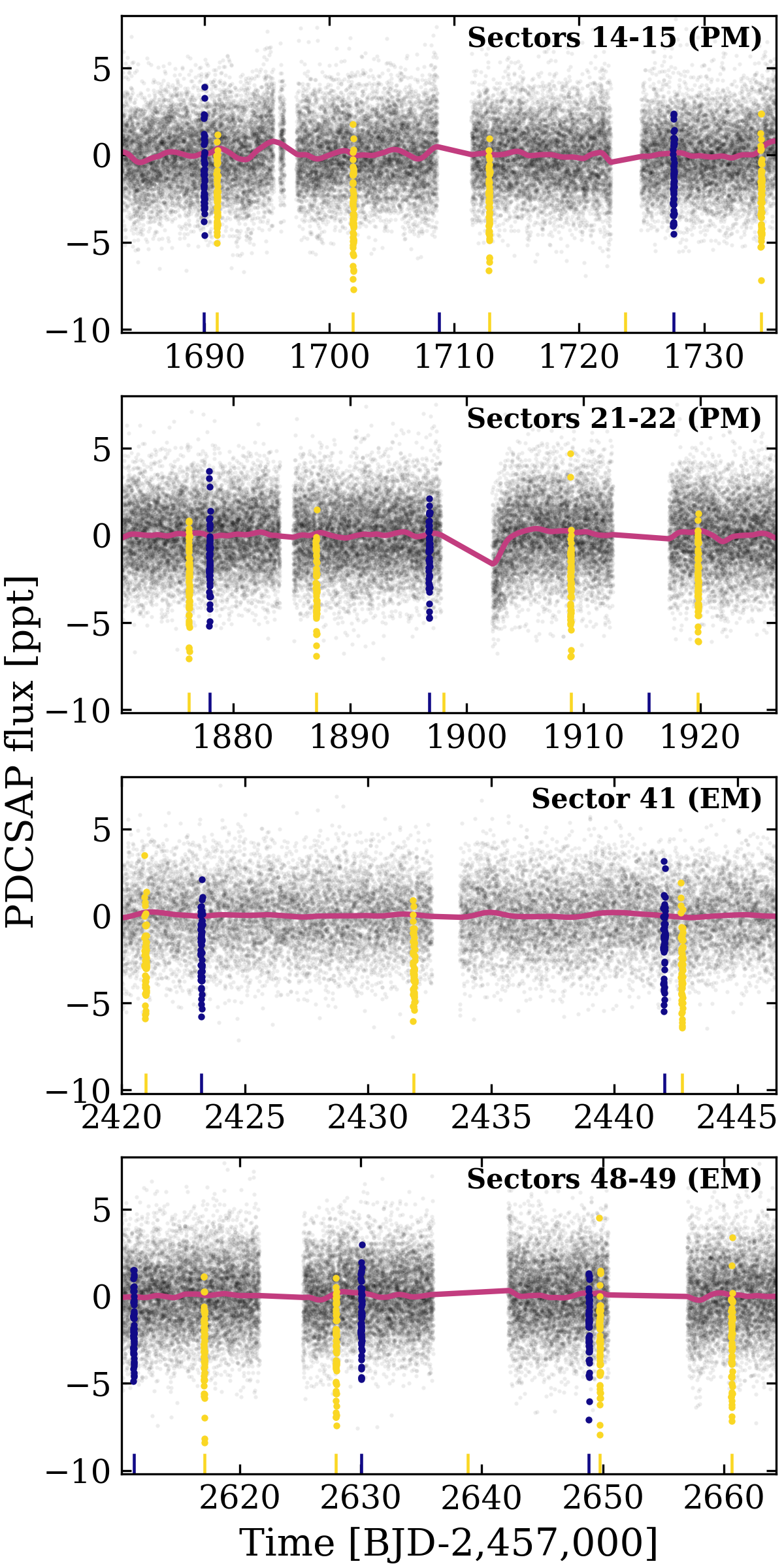}
    \caption{TESS PDCSAP light curves of TOI-1266 from the primary mission (PM) and first extended mission (EM). The mean GP detrending model for each set of consecutive sectors is shown in pink. The bold measurements and vertical ticks highlight the transit times of the planets TOI-1266 b (yellow) and c (blue), respectively.}
    \label{fig:pdcsap}
\end{figure}

\section{Observations} \label{sect:obs}
\subsection{TESS transit photometry} \label{sect:obs:tess}
\begin{figure*}
    \centering
    \includegraphics[width=\hsize]{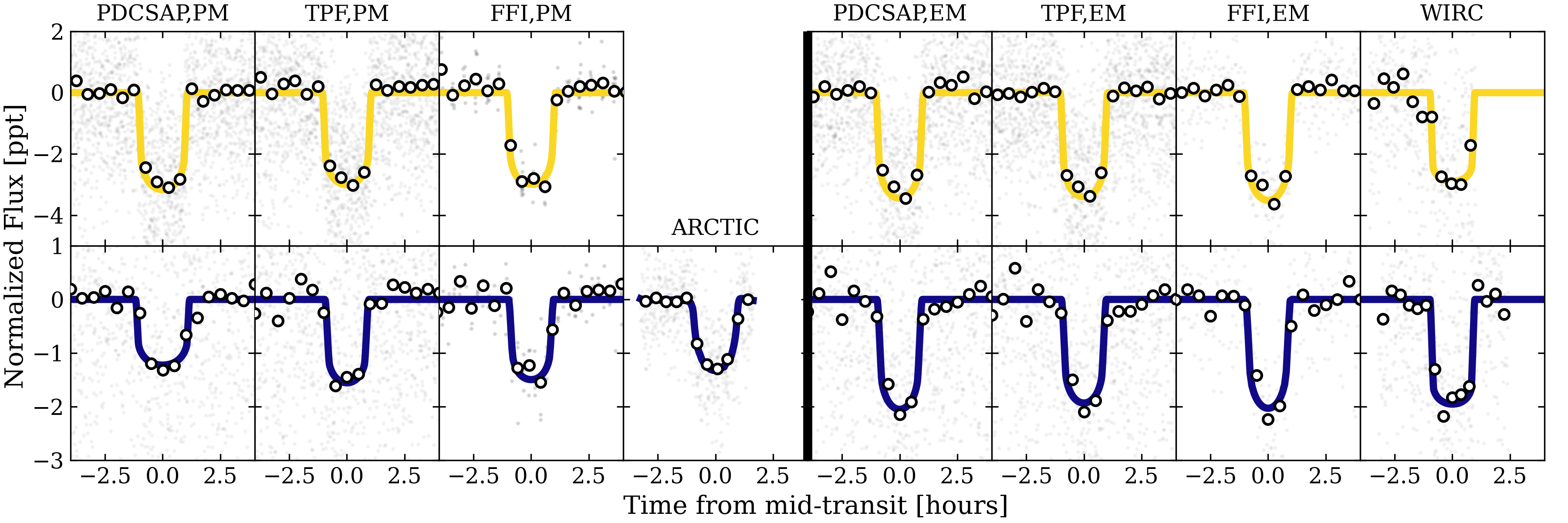}
    \caption{Phase-folded transits and transit model fits of both TOI-1266 planets from three TESS data reductions and two ground-based facilities. The transits of TOI-1266 b and c are depicted in the top and bottom rows, respectively. Different TESS extractions from the primary (PM) and extended missions (EM) include PDCSAP and our custom extractions from the TESS TPFs and FFIs. We also include ground-based, diffuser-assisted transit observations from the APO/ARCTIC (TOI-1266 c only) and Palomar/WIRC cameras, which we use to validate the variations in both planets' transit shapes revealed by TESS between the PM and EM. The central vertical bar separates observations taken during the PM on the left and post-PM on the right (i.e. on and after UT 2021 July 24).}
    \label{fig:transits}
\end{figure*}

TOI-1266 b and c were originally discovered in the primary mission (PM) of NASA's Transiting Exoplanet Survey Satellite \citep[TESS;][]{Ricker_2015}, as reported by \citetalias{Demory_2020} and \citetalias{Stefansson_2020}. Both discovery papers presented data from TESS sectors 14, 15, 21, and 22, which spanned UT 2019 July 18 to 2020 March 17. TOI-1266 has since been reobserved in the TESS's first extended mission (EM) in sectors 41, 48, and 49 (UT 2021 July 24 to 2022 March 25). The aforementioned discovery papers focused their TESS analyses on the 2-minute Presearch Data Conditioning Simple Aperture Photometry (\texttt{PDCSAP}) light curves \citep[PDCSAP;][]{twicken10, morris20}, which were processed by the NASA Ames Science Processing Operations Center \citep[SPOC;][]{jenkins16}. The full PDCSAP light curve is depicted in Figure~\ref{fig:pdcsap} and the phase-folded transit light curves from the PM and EM are included in Figure~\ref{fig:transits}.

In this paper we use the PDCSAP light curves in addition to custom light curve extractions from the 2-minute TESS Target Pixel Files (TPFs) and the Full Frame Images (FFIs). The latter has cadences of 30 and 10 minutes in the TESS PM (i.e. sectors 14,15,21,22) and EM (i.e. sectors 41,48,49), respectively. We consider multiple light curve extractions to assess the validity of each planet's transit shape across all seven TESS sectors (see Section~\ref{sect:analysis:transit}). We performed a custom extraction of the TESS 2-minute light curve from the TPFs using the \texttt{lightkurve} package \citep{lightkurve_2018,astropy_2022,astroquery_2019}. Similarly to the PDCSAP light curve extraction, we used the cotrending basis vectors (CBVs) for the appropriate TESS camera and CCD in each sector to correct for systematics. This allows us to have more control over the light curve extraction to assess the sensitivity of the planetary transit depths on the extraction methodology (see Section~\ref{sect:analysis:transit}). We tested a variety of photometric apertures including the optimized TESS aperture size and threshold apertures that include pixels within 3, 5, and 10$\sigma$ of the median target flux. We also corrected for dilution by neighbouring sources based on their known positions and TESS magnitudes in the TESS Input Catalog \citep[TIC;][]{Stassun_2019}. We derived dilution values that are consistent with the values reported by the SPOC and used in the PDCSAP analysis. The phase-folded TPF transit light curves for both planets from the PM and EM are included in Figure~\ref{fig:transits}.  

We performed a second custom light curve extraction from the TESS FFIs following \cite{Vanderburg_2019}. We retrieved the TPFs using the TESSCut interface integrated into MAST \citep{Brasseur_2019}. We extracted each sector's light curve using 20 photometric apertures that were either circularly or PSF-shaped. In each sector, we selected the aperture size and shape that maximized the photometric precision after correcting for dilution and detrending via linear regression using the spacecraft's quaternion time series and the PDCSAP CBVs, after binning to the appropriate FFI cadence (i.e. 30 minutes in the PM and 10 minutes in the EM). An additional spline component was included to correct for any residual photometric variability that we attribute to the star. The phase-folded FFI transit light curves for both planets from the PM and EM are included in Figure~\ref{fig:transits}.

A preliminary inspection of transit light curves in Figure~\ref{fig:transits} revealed significant transit depth discrepancies between the PM and EM. We postpone our investigation of these anomalies until Section~\ref{sect:analysis:transit}.

\subsection{APO/ARCTIC diffuser-assisted photometry} \label{sect:arctic}
Figure~\ref{fig:transits} includes an archival ground-based transit of TOI-1266 c taken on UT 2020 January 28 with the diffuser-assisted ARCTIC camera on the 3.5-m ARC telescope at Apache Point Observatory (APO). These observations were originally presented by  \citetalias{Stefansson_2020}, with the details described therein.

\subsection{Palomar/WIRC diffuser-assisted photometry} \label{sect:wirc}
We observed one full transit of each of TOI-1226 b and c in the $J$-band with the Wide-field Infared Camera (WIRC) on the Hale Telescope at Palomar Observatory, California, USA. Palomar/WIRC is a 5.08-m telescope equipped with a $2048\times 2048$ Rockwell Hawaii-II NIR detector, providing a field of view of 
8\farcm7 $\times$ 8\farcm7 with a plate scale of 0\farcs25 per pixel \citep{Wilson_2003}. Our data were taken with a beam-shaping diffuser that increased our observing efficiency and improved the photometric precision and guiding stability \citep{Stefansson_2017,Vissapragada_2020}. 

We observed the transit of TOI-1266 c on UT 2022 March 10 between airmasses of 1.65-1.19. Our observing window provided close to three full transit durations of coverage, centered on the transit mid-point. We collected 660 science images with total exposure times of 10.5 seconds per image. We observed the transit of TOI-1266 b on UT 2022 June 6 between airmasses of 1.26-2.26. Our observing window provided pre-ingress coverage and full transit coverage, but no post-egress baseline.  

\begin{figure*}
    \centering
    \includegraphics[width=\hsize]{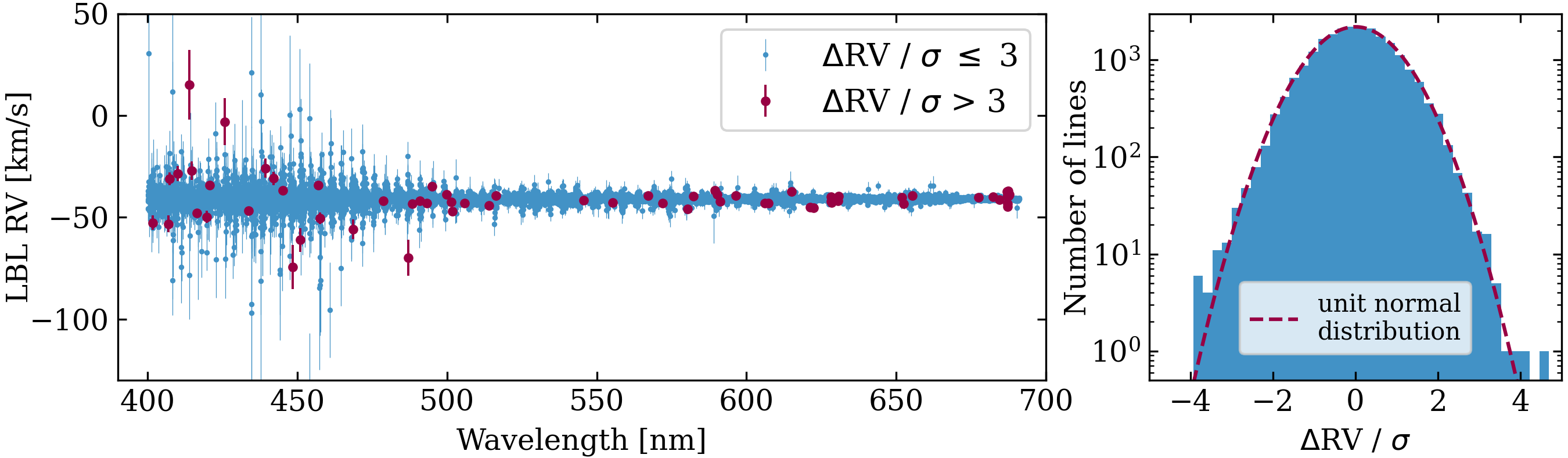}
    \caption{An illustrative example of our line-by-line (LBL) RV extraction from a representative HARPS-N spectrum. \emph{Left panel}: the measured RV and uncertainty of 22,577 individual lines (light blue) as a function of wavelength. The dark purple markers depict 48 lines that are rejected as outliers with $\Delta \mathrm{RV} / \sigma > 3$ in this spectrum, where $\Delta RV$ represents the RV differences of individual lines from the median value, and $\sigma$ are the RV uncertainties on individual lines. \emph{Right panel}: the distribution of normalized LBL RVs compared to the unit normal distribution (dashed purple line).}
    \label{fig:lbl}
\end{figure*}

We collected 517 science images with the same total exposure time as our first transit. For both observing sequences, we initiated the observations with a nine-point dither near the target to construct a background frame that we used to correct for the effects of a known detector systematic at short exposure times. We dark-subtracted, flat-fielded, and corrected for bad pixels following the methodology of \cite{Vissapragada_2020}. We scaled and subtracted the aforementioned background frame from each image to remove the full frame background structure. We then performed circular aperture photometry with the \texttt{photutils} package \citep{Bradley_2020} on our target, plus three comparison stars. We tested apertures sizes between 7-25 pixels and selected the aperture that minimized the the root-mean-square deviation of the final photometry. We also used uncontaminated annuli with inner and outer radii of 25 and 35 pixels, respective, for local background subtraction. We fit the light curves using the \texttt{exoplanet} package \citep{foremanmackey19,exoplanet:joss}, which included a systematics correction constructed as a linear combination of comparison star light curves, the centroid offset, PSF width, airmass, and local background 
\citep{Greklek-McKeon_2023}. Typically, airmass and sky background are not included simultaneously as detrending parameters, but their inclusion here provided improved fits as our observations showed considerable variations in background flux that did not track with airmass, despite the clear weather on both nights. Our light curves are included in the right-most column of Figure~\ref{fig:transits}.

\subsection{Precise radial velocity measurements from HARPS-N} \label{sect:harpsn}
We obtained \nrv{} spectra of TOI-1266 with the HARPS-N spectrograph between UT 2020 May 8 and 2022 July 14. HARPS-N is a high resolution ($R=115,000$) optical \'{e}chelle spectrograph located at the 3.6m Telescopio Nazionale Galileo (TNG) on La Palma, Canary Islands. HARPS-N boasts exceptional long-term pressure and temperature stability, which enables the instrument to reach sub-m/s stability on bright FGK stars \citep{cosentino12}. Despite its operation at optical wavelengths (i.e. 380-691 nm), HARPS-N has consistently demonstrated its ability to reach the \mps{} level of precision on bright ($V\lesssim 13.5$) early-to-mid M dwarfs (LTT 3780;~\citealt{Cloutier_2020a}, TOI-1235;~\citealt{Cloutier_2020b}, TOI-1634;~\citealt{Cloutier_2021}, TOI-1695;~\citealt{Cherubim_2023},  HD 79211;~\citealt{DiTomasso_2023}, LHS 1903; Wilson et al. submitted).

Each HARPS-N spectrum was taken as part of the HARPS-N Guaranteed Time Observations (GTO) program.
We positioned the non-science fibre on-sky in all of our observations.
We fixed all exposure times to 1800 s. We reduced our spectra using the latest version of the HARPS-N Data Reduction Software v2.3.5 \citep[DRS;][]{Dumusque_2021}. We obtain a nightly RV measurement from the DRS via the cross-correlation function (CCF) technique using an M2 CCF mask. The DRS also produces time series of the full width at half maximum (FWHM) and bisector inverse slope (BIS) activity indicators. We obtained a median peak signal-to-noise ratio (S/N) across our spectra of 33.

We performed our RV extraction using the novel line-by-line (LBL) method proposed by \cite{Dumusque_2018} and implemented by \cite{Artigau_2022}. We used version 0.61.0 of the \texttt{LBL} code\footnote{\url{https://github.com/njcuk9999/lbl}}, which performs a simple telluric correction by fitting a TAPAS model \citep{Bertaux_2014}. This routine has been demonstrated to significantly improve the RV precision of M dwarfs observed with ESPRESSO \citep{Allart_2022}.

The LBL method is conceptually similar to the commonly used template-matching method, which makes LBL similarly well-suited to RV extraction from M dwarf spectra \citep{anglada12,astudillodefru15}. It is a data-driven method wherein the Doppler shifts of many individual spectral lines are calculated with respect to a high S/N template spectrum that we construct by coadding all our HARPS-N spectra. LBL extractions have demonstrated similar performances to template-matching in terms of RV precision and dispersion on M dwarfs observed with optical RV spectrographs \citep[e.g.][]{Artigau_2022}. We elect to use an LBL extraction over template-matching because the former is robust to outlying RV measurements that can bias a spectrum's RV measurement well beyond the typical photon noise uncertainty. Spurious lines can arise from a variety of sources including imperfect telluric contamination, detector defects, and variable line sensitivity to magnetic activity \citep[e.g.][]{Lafarga_2023}. Spurious and high S/N signals offset a spectrum's RV measurement when left unaccounted for, which is the case for template-matching extractions that consider all lines. 

By calculating the RV shift of individual lines, we easily identify high S/N outliers and reject them, therefore producing a more robust RV measurement. In each spectrum, we reject lines that are $>3\sigma$ discrepant from the median RV over all lines. An illustrative example of this is shown in Figure~\ref{fig:lbl} for a randomly selected HARPS-N spectrum.
We perform our outlier rejection on the full HARPS-N wavelength domain, as well as in the $BVR$-bands, which we will ultimately use to assess the chromaticity of RV signals (Section~\ref{sect:analysis:gls}). The rms values and median uncertainties in each RV time series, including the CCF RVs, are reported in Table~\ref{tab:chrom}. Indeed, LBL offers a significant improvement over the CCF technique for TOI-1266.

\begin{table}
  \centering
  \caption{Summary of RV time series \label{tab:chrom}}
  \begin{tabular}{lccc}
    \hline
    RV time series & Wavelength range & RMS & Median uncertainty \\
    & [nm] & [\mps{]} & [\mps{]} \\
    \hline
    Full LBL & 380-691 & 3.74 & 1.91 \\
    $B$-band LBL & 380-492 & 19.85 & 18.29 \\
    $V$-band LBL & 507-595 & 6.71 & 5.32 \\
    $R$-band LBL & 589-691 & 4.62 & 4.52 \\
    Full CCF & 380-691 & 5.52 & 4.54 \\
    \hline
  \end{tabular}
\end{table}

We also calculated the differential line width activity indicator dLW using the LBL framework \citep{Zechmeister_2018,Artigau_2022}. The dLW time series has units of m$^2$/s$^2$ and scales with the FWHM of the line profile. But similarly to our LBL RVs, the dLWs are constructed in an outlier-resistant way. Because of this, and because of the generally poor performance of M dwarf binary masks in the CCF method, we consider the dLW time series to be a more robust activity indicator than the CCF FWHM and BIS. We provide the dLW, along with all of our spectroscopic LBL and CCF time series, in Table~\ref{tab:rv}.

\section{TOI-1266 Stellar Parameters} \label{sect:star}
TOI-1266 has been studied extensively in previous efforts to validate the planetary system (\citetalias{Demory_2020}, \citetalias{Stefansson_2020}). Spectroscopic reconnaissance, high-resolution imaging, and Gaia astrometry \citep[RUWE=1.227;][]{GaiaDR3} rule out comoving stellar companions and indicate that TOI-1266 is a single star located at a distance of 36 pc.

TOI-1266 is considered magnetically inactive and does not show signs of photometric variability in its TESS light curve (Figure~\ref{fig:pdcsap}). We are unable to recover the stellar rotation period in the TESS data. We also queried the ground-based photometric monitoring campaigns from the All-Sky Automated Survey for SuperNovae \citep[ASAS-SN;][]{Shappee_2014,Kochanek_2017} and the Zwicky Transient Facility \citep[ZTF;][]{ZTF_2019}. We find no signatures of stellar rotation in these photometric light curves but we do find evidence for \prot{} $\sim 45$ days in our spectroscopic time series, presumably from bright plages (see Section~\ref{sect:analysis:gls}). The lack of rotational modulation in the TESS, ASAS-SN, and ZTF light curves indicates that TOI-1266's filling factor by dark starspots is consistent with zero. We also measure \logrhk{}$=-5.50^{+0.35}_{-0.44}$ from the Mount Wilson S-index calculated by the HARPS-N DRS \citep{Astudillo_2017}, which supports the notion that TOI-1266 is chromospherically inactive.

We adopt physical stellar parameters from the TESS Input Catalog v8.2 \citep[TIC;][]{Stassun_2019}, which are consistent with the values reported in \citetalias{Demory_2020} and \citetalias{Stefansson_2020}. Values of the stellar effective temperature, radius, and mass are based on the $G_{\mathrm{BP}}$, $G_{\mathrm{RP}}$, and 2MASS $K_s$-band magnitudes and use the empirical relations for M dwarfs from \cite{Mann_2013,Mann_2015,Mann_2019}, respectively. The TIC does not report a metallicity measurement but we note that both discovery papers present multiple [Fe/H] measurements using spectroscopic and SED-fitting techniques. The five literature values of [Fe/H] exhibit a large rms of 0.17 dex and we do not attempt to evaluate the accuracy of these measurements. Instead, we conclude that TOI-1266 appears to be somewhat metal-poor and report a range of metallicity values in Table~\ref{tab:star}, along with all other stellar parameters.

\begin{table}
  \centering
  \caption{TOI-1266 stellar parameters \label{tab:star}}
  \begin{tabular}{lcc}
    \hline
    \multicolumn{3}{c}{\emph{TOI-1266, TIC 467179528, 2MASS J13115955+6550017,}} \\
    \multicolumn{3}{c}{\emph{Gaia DR3 1678074272650459008}} \\
    \hline
    \multicolumn{3}{c}{\emph{Astrometry}} \\
    Right ascension (J2016.0), $\alpha$ & 13:11:59.1 & 1 \\ 
    Declination (J2016.0), $\delta$ & +65:50:1.3 & 1 \\ 
    RA proper motion,  & $-150.557\pm 0.014$ & 1 \\
    $\mu_{\alpha}$ [mas yr$^{-1}$] && \\
    Dec proper motion, & $-25.339\pm 0.013$ & 1 \\
    $\mu_{\delta}$ [mas yr$^{-1}$] && \\
    Parallax, $\pi$ [mas] & $27.740\pm 0.013$ & 1 \\
    Distance, $d$ [pc] & $36.049\pm 0.016$ & 2 \\
    RUWE & $1.227$ & 1 \\
    \multicolumn{3}{c}{\emph{Photometry}} \\
    NUV$_{\text{GALEX}}$ & $23.07 \pm 0.55$ & 3 \\ 
    $V$ & $12.941\pm 0.049$ & 4 \\
    $G_{\mathrm{BP}}$ & $13.2478\pm 0.0010$ & 1 \\
    $G$ & $12.1222\pm 0.0003$ & 1 \\
    $G_{\mathrm{RP}}$ & $11.5011\pm 0.0007$ & 1 \\
    $T$ & $11.040 \pm 0.007$ & 5 \\
    $J$ & $9.706\pm 0.023$ & 6 \\ 
    $H$ & $9.065\pm 0.030$ & 6 \\ 
    $K_s$ & $8.840\pm 0.020$ & 6 \\
    \multicolumn{3}{c}{\emph{Stellar Parameters}} \\
    Spectral type & M3V & 7 \\
    $M_{K_s}$ & $6.624\pm 0.020$ & 8 \\
    Surface gravity, \logg{} [dex] & $4.793^{+0.032}_{-0.033}$ & 8 \\
    Effective temperature, & $3618\pm 157$ & 5 \\
    \teff{} [K] && \\
    Metallicity, [Fe/H] [dex] & $[-0.31,-0.08]$ & 9,10 \\
    Stellar radius, $R_\star$ [$\mathrm{R}_{\odot}$] & $0.436\pm 0.013$ & 5 \\ 
    Stellar mass, $M_\star$ [$\mathrm{M}_{\odot}$] & $0.431\pm 0.020$ & 5 \\
    Stellar density, $\rho_\star$ [g cm$^{-3}$] & $7.33^{+0.76}_{-0.71}$ & 8 \\
    Stellar luminosity, $L_\star$ [$\mathrm{L}_{\odot}$] & $0.029^{+0.006}_{-0.005}$ & 8 \\
    \vspace{-0.15cm} Projected rotation velocity, && \\ \vspace{-0.25cm}
    & $<1.3$ & 8 \\
    \vsini{} [km s$^{-1}$] && \\
    Rotation period, \prot{} [days] & $44.6^{+0.5}_{-0.8}$ & 8 \\
    \logrhk{} & $-5.50^{+0.35}_{-0.44}$ & 8 \\
    \hline
    \multicolumn{3}{l}{\textbf{References:} 1) \cite{GaiaDR3} 2) \cite{Bailer_2021}} \\ 
    \multicolumn{3}{l}{3) \cite{bianchi17} 4) \cite{zacharias13} 5) \cite{Stassun_2019}} \\
    \multicolumn{3}{l}{6) \cite{cutri03} 7) \cite{pecaut13} 8) this work 9) \citetalias{Demory_2020} 10) \citetalias{Stefansson_2020}.}
  \end{tabular}
\end{table}

\section{Data Analyses} \label{sect:analysis}
\subsection{Generalized Lomb-Scargle periodograms} \label{sect:analysis:gls}
\begin{figure*}
  \centering
  \includegraphics[width=0.85\hsize]{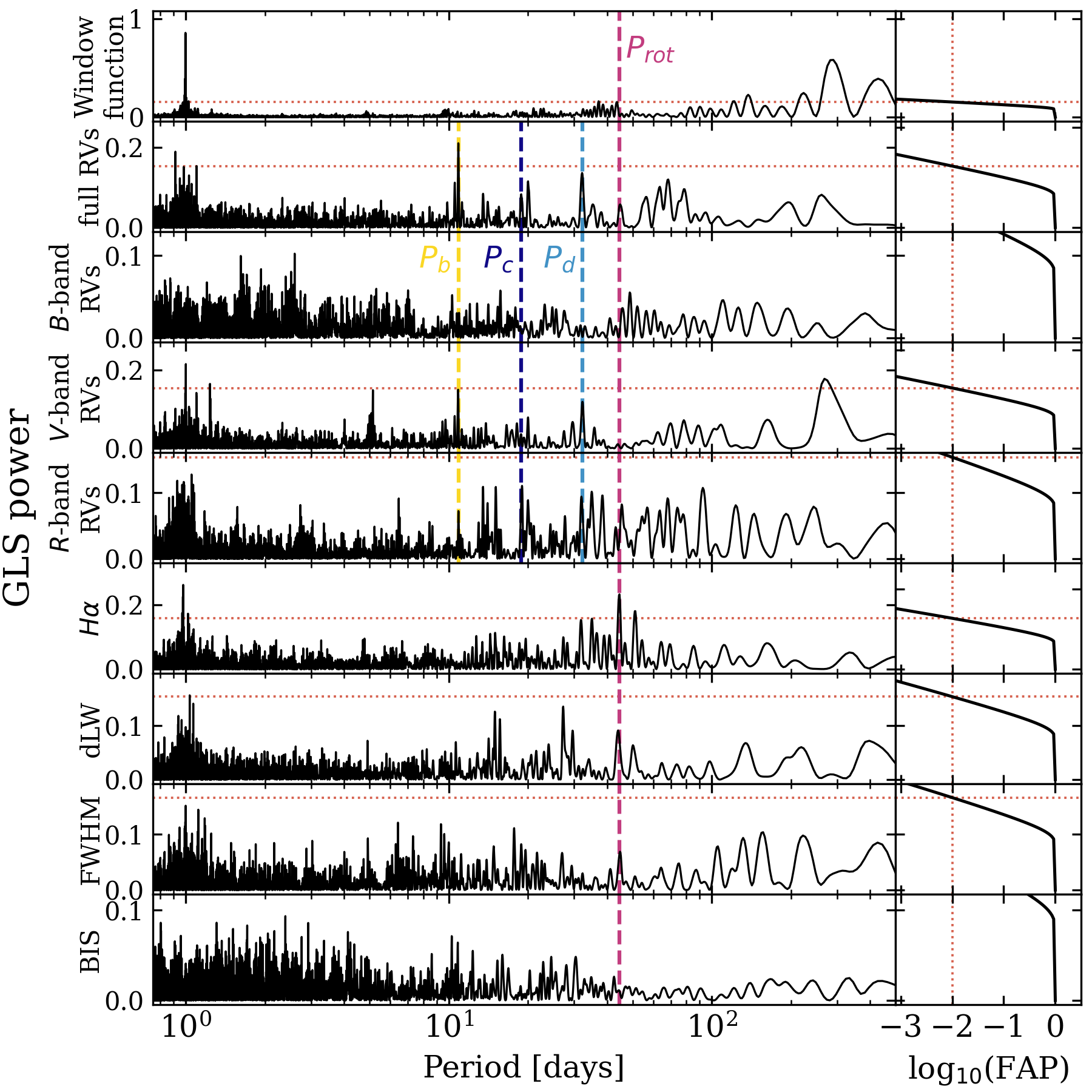}
  \caption{GLS periodograms of our HARPS-N window function, the full LBL RVs, the chromatic LBL RVs in the $BVR$-bands, and the spectroscopic activity indicators $H\alpha$, dLW, FWHM, and BIS. Left column: the GLS periodogram of the time series labeled on the y-axis. The vertical dashed lines highlight the known orbital periods of the transiting planets TOI-1266 b and c, along with the prominent peaks at 32.3 and 44.6 days in the full RVs and $H\alpha$ time series that we attribute to a new planetary candidate and to stellar rotation, respectively. The planetary and candidate orbital periods are only depicted in the four RV panels for clarity. The horizontal dotted lines depict each periodogram's 1\% FAP level. Right column: the FAP as a function of GLS power corresponding to each periodogram.}
  \label{fig:gls}
\end{figure*}

Figure~\ref{fig:gls} depicts the generalized Lomb-Scargle periodograms \citep[GLS;][]{zechmeister09} of our spectroscopic window function, the full RVs, the chromatic $BVR$ RVs, and the following activity indicators: $H\alpha$, dLW, FWHM, and BIS. The corresponding false alarm probability (FAP) of each time series is included in Figure~\ref{fig:gls}.

Aside from the aliased signals close to 1-day \citep{Dawson_2010}, the GLS of the full RV time series exhibits one periodicity with FAP $< 1$\% at 10.9 days, which corresponds to the transiting planet TOI-1266 b. Other noteworthy peaks are seen at the orbital period of TOI-1266 c (i.e. 18.8 days; FAP $=95$\%), plus a peak at 32.3 days that is unrelated to either known transiting planet (FAP $=4$\%). We investigated the origin of the latter by assessing its chromatic dependence in the $BVR$-bands and its presence in the GLSs of various activity indicators. The chromatic RVs are not informative because the $B$-band RVs are too low S/N, and while the 32.3-day signal does exist in the $V$ and $R$-band RVs with varying FAP values, so do the known signals from the transiting planets. In particular, the prominent signal from TOI-1266 b decreases from the $V$-band to the $R$-band RVs, which is not expected as the relative power of achromatic planetary signals are expected to increase with increasing wavelength as the impact of magnetic activity is lessened \citep{Reiners_2010}. Establishing the chromaticity of GLS signals is clearly questionable from our time series. We therefore refrain from commenting on the origin of 32.3-day signal until Section~\ref{sect:evidence}. 

Turning to the activity indicators, we do not see a low FAP peak close to 32.3 days in any of our activity time series. The dLW, FWHM, and BIS time series do not show any signs of stellar activity and are all consistent with noise. The only prominent activity peak with FAP $< 1$\% is exhibited by our $H\alpha$ time series at 44.6 days. $H\alpha$ is a strong indicator of M dwarf chromospheric activity at optical wavelengths \citep{Lafarga_2021} and we interpret the 44.6-day signal as the star's rotation period, or a harmonic thereof. The lack of a low FAP peak at 32.3 days, or $32.3/2 \approx 16.2$ days, in any of the activity indicators leads to the possibility that this periodicity may be due to a third planet around TOI-1266. We reserve the justification of this claim until Section~\ref{sect:evidence} wherein we compare the model evidences for two planets versus three planets based on our RV data.

\subsection{Global RV + Transit Modeling} \label{sect:analysis:global}
Here we jointly model our HARPS-N RVs and the TESS PDCSAP light curves. We consider the TESS PM and EM data independently in order to separate the distinct transit depths between each dataset.

We set the planetary components of our global model to be pseudo-Keplerian orbits. We consider pseudo-Keplerians because the period ratio of TOI-1266 b and c is within 3.5\% of the 5:3 mean motion resonance (MMR) and \citetalias{Demory_2020} showed marginal evidence for transit timing variations (TTVs) of both planets in the TESS PM. We note however that the independent analysis of the same TESS data by \citetalias{Stefansson_2020} concluded that the system does not exhibit measurable TTVs. Here we assume the general case in which TTVs may be present and measurable. 

The maximum TTV amplitudes reported by \citetalias{Demory_2020} are 4 minutes and 16 minutes for planets b and c, respectively, but these measurements represent only marginal TTV detections at $< 3\sigma$. The corresponding TTV masses are poorly constrained: $M_{p,b}=13.5^{+11.0}_{-9.0}\, \mathrm{M}_{\oplus}$ and $M_{p,c}=2.2^{+2.0}_{-1.5}\, \mathrm{M}_{\oplus}$ \citepalias{Demory_2020}. Using the WHFast integrator in the \texttt{rebound} N-body code \citep{Rein_2012,Rein_2015}, we estimate that non-Keplerian effects result in RV excursions from Keplerian motions at the level of $\sim 30$ cm/s. These excursions are $\sim 2-3\times$ smaller than our forthcoming precision on the planets' semiamplitudes (i.e. $\sim 70-90$ cm/s) and do not correlate with our uniformly sampled RV observations such that they are a negligible effect. The planet-induced transit and RV signals that we aim to extract from our data are well-approximated as being Keplerian, although our pseudo-Keplerian model retains the flexibility to measure TTVs. This is accomplished by fixing each planets' orbital elements while allowing individual mid-transit times to vary. This makes us sensitive to TTVs but not to transit duration variations.

The light curve component of our global model features the following parameters: stellar mass and radius, flux baseline $f_0$, scalar jitter $s$, quadratic limb-darkening parameters $\{u_1,u_2\}$, orbital periods $P$, planet-star radius ratios $R_p/R_\star$, impact parameters $b$, and rescaled orbital eccentricities $e$ and arguments of periastron $\omega$ (i.e. $\{ \sqrt{e}\cos{\omega},\sqrt{e}\sin{\omega} \}$). The RV component includes the additional parameters: the systemic RV $\gamma_0$, scalar jitter $s_{\rm RV}$, and each planet's RV semiamplitude $K$. Our adopted priors are summarized in Table~\ref{tab:results}.

We consider two sets of models containing two and three planets, respectively. The latter set of models include a third Keplerian component to model the periodic signal at 32.3 days of insofar unknown origin (c.f. Figure~\ref{fig:gls}). Additionally, we consider separate models that do and do not include a Gaussian process (GP) regression model of temporal correlations from stellar activity in our RV time series. We elect to train the GP on our $H\alpha$ time series as it exhibits the periodic signal at $\sim 44.6$ days that we attribute to stellar rotation, and because GP training on photometry is not informative when photometric signatures of activity are undetected \citep{Tran_2023}. The simultaneity of the $H\alpha$ and RV time series also has the desirable effect of making us insensitive to temporal changes in the RV activity signal.

We use the \texttt{exoplanet} software package \citep{foremanmackey19,exoplanet:joss} to build our RV+transit model, which uses the \texttt{celerite2} package \citep{celerite1,celerite2} to construct the GP and evaluate the likelihood of our dataset under the GP model. The \texttt{starry} package is used within \texttt{exoplanet} to construct our model light curve \citep{Luger_2019}. For the RV component of our global model, we adopt a covariance kernel whose power spectral density is the superposition of two damped simple harmonic oscillators (SHO). The SHO's damping factor produces a covariance kernel that is qualitatively similar to the commonly used quasi-periodic kernel. Similarly, we adopt a GP model of residual systematics in the TESS light curves using a kernel with one damped SHO, whose hyperparameter values are distinct from the RV component. We refrain from describing the details of our covariance kernels as they have been described extensively throughout the literature \citep[e.g.][]{Cherubim_2023} and in the \texttt{celerite2} documentation\footnote{\url{https://celerite2.readthedocs.io/en/latest/api/python/\#celerite2.terms.RotationTerm}}. Our GP hyperparameter priors are included in Table~\ref{tab:results}.

We use the \texttt{PyMC3} Markov Chain Monte Carlo (MCMC) package
\citep{salvatier16} to sample the joint posterior probability density function (PDF) of our global model parameters. The MCMC is initialized with two simultaneous chains, each with $10^3$ burn-in steps and $10^4$ posterior draws. We report maximum a-posteriori (MAP) point estimates from each parameter's marginalized posterior PDF in Table~\ref{tab:results}. Parameter uncertainties are derived from each marginalized posterior's 16$^{\rm th}$ and 84$^{\rm th}$ percentiles unless noted otherwise. 

\begin{figure*}
    \centering
    \includegraphics[width=0.92\hsize]{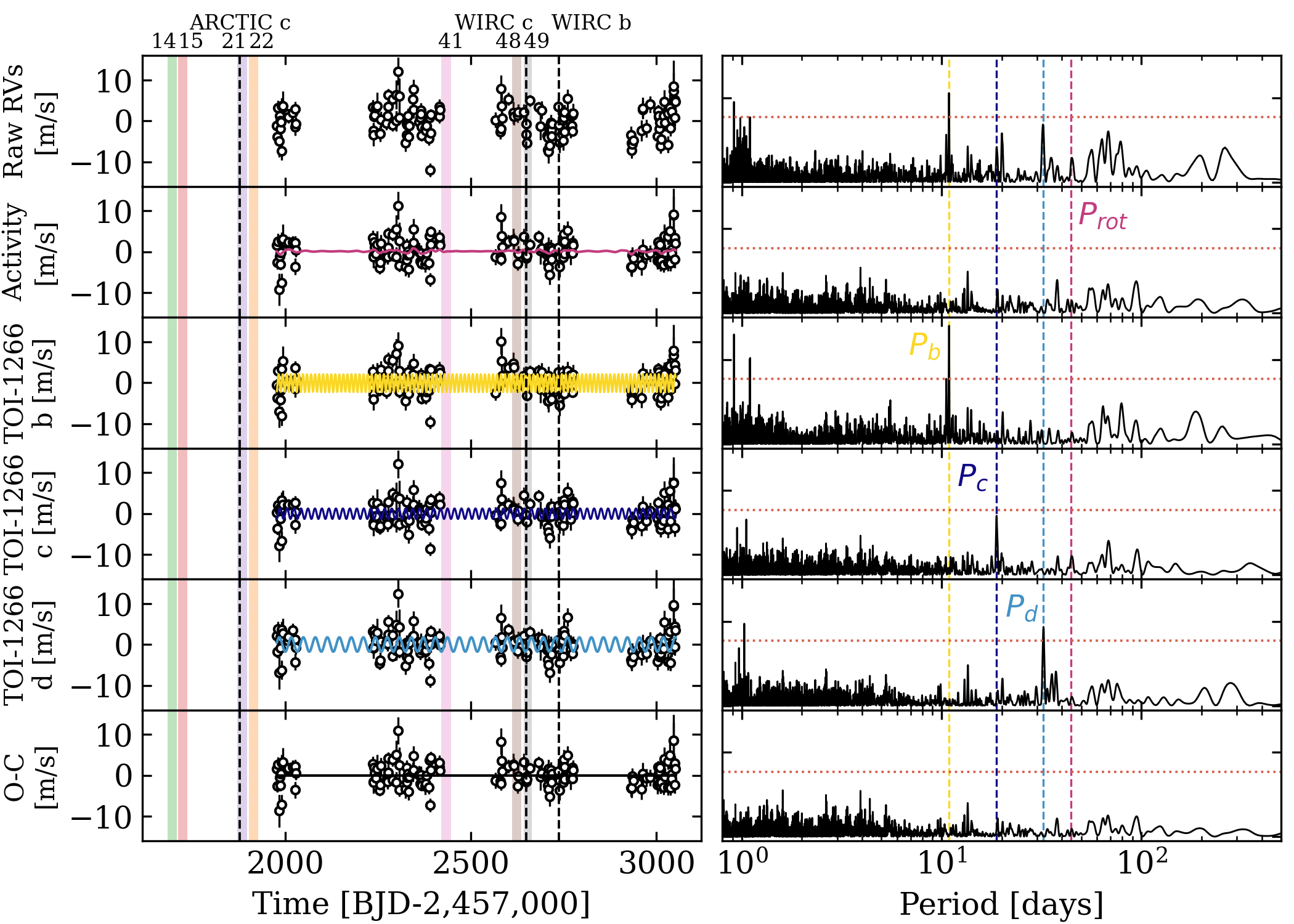}
    \caption{Individual components of our 3-planet RV model of TOI-1266 and corresponding GLS periodograms. From top to bottom: the raw LBL RVs, stellar RV activity, TOI-1266 b, TOI-1266 c, the planet candidate TOI-1266 d, and the residuals. The residuals do not show any evidence for unaccounted periodicities in the GLS. The vertical shaded regions in the left column highlight each TESS sector's observing window (annotated at the top) and reveals where we have contemporaneous TESS and spectroscopic observations. Similarly, the vertical dashed lines highlight the epochs of our ground-based transit observations of either TOI-1266 b or c, using the ARCTIC and WIRC cameras. The vertical dashed lines in the right column depict the orbital periods of the transiting TOI-1266 planets, the candidate planet d, and the stellar rotation period inferred from our $H\alpha$ time series.}
    \label{fig:rvs}
\end{figure*}

\begin{figure*}
    \centering
    \includegraphics[width=0.99\hsize]{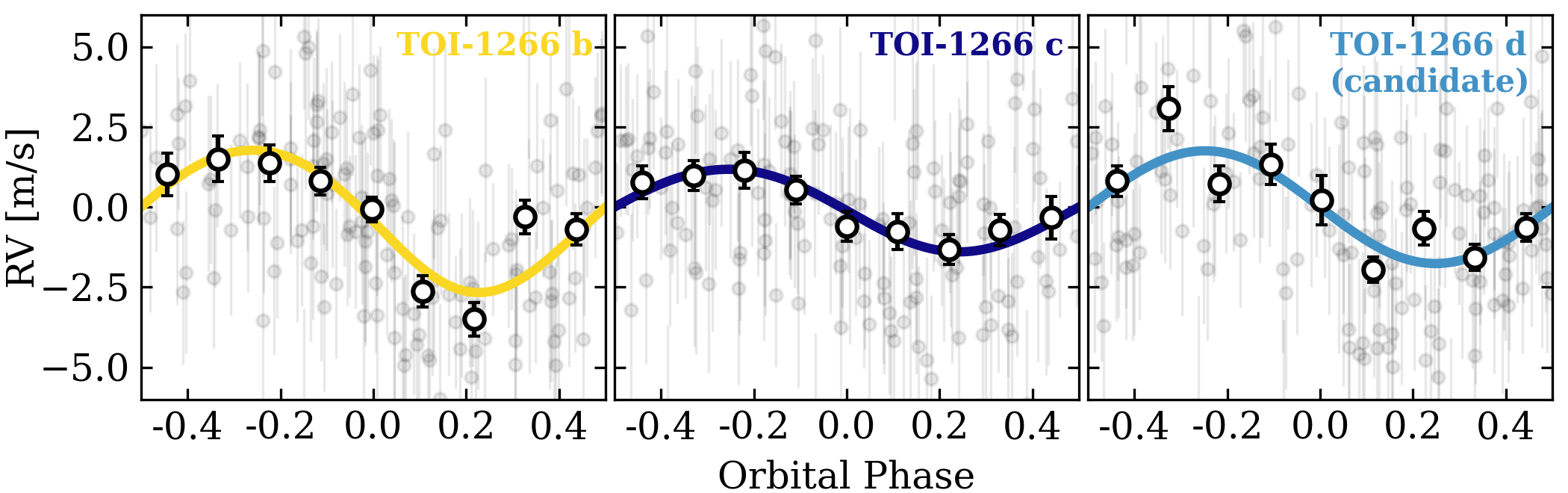}
    \caption{Phase-folded RV time series for the transiting planets TOI-1266 b (left) and c (middle), and the planet candidate TOI-1266 d (right).}
    \label{fig:phased}
\end{figure*}

The detrended transit light curves were shown in Figure~\ref{fig:transits} while the individual RV components are in Figure~\ref{fig:rvs} for our fiducial model containing three Keplerian components and a GP. The GLS periodogram of each Keplerian component reveals that the Keplerian's period is the lowest FAP signal, while the stellar rotation period of 44.6 days is not present in the GLS of the RV activity component. However, in Section~\ref{sect:evidence} we will show that our fiducial model, which includes a GP, is favoured. The GLS of the RV residuals shows no evidence for significant residual periodicities. Figure~\ref{fig:phased} depicts the phase-folded RV time series for each planet.

\begin{table}
  \centering
  \caption{Bayesian evidences for the four RV + TTV models considered. \label{tab:lnZ}}
  \begin{tabular}{cccc}
    \hline
    Model, $M$ & $\ln{\mathcal{Z}_M}$ & Bayes factor$^*$ & Bayes factor$^*$ \\
    && $\log_{10}{(\mathcal{Z}_M/\mathcal{Z}_{2p})}$ & $\log_{10}{(\mathcal{Z}_{M}/\mathcal{Z}_{2p+GP})}$ \\
    \hline
    2p & -398.3 & 0 & $-3.15$ \\
    2p + GP & -391.0 & 3.15 & 0 \\
    3p & -385.0 & 5.6 & 2.4  \\
    3p + GP & - 375.8 & 9.6 & 6.4  \\
    \hline    
    \multicolumn{4}{l}{$^*$ Bayes factors are scaled by the ratio of model priors $p(N_{\mathrm{pl}}) = (1/3)^{N_{\mathrm{pl}}}$,} \\
    \multicolumn{4}{l}{where $N_{\mathrm{pl}}$ is the number of planets in the model \citep{Nelson_2020}.} \\
  \end{tabular}
\end{table}

\begin{table*}
  \centering
  \caption{Priors and posterior point estimates of the TOI-1266 planetary system model parameters.\label{tab:results}}
  \begin{tabular}{lcccc}
    \hline
    Parameter & Prior & \multicolumn{3}{c}{Posterior Point Estimates} \\
    \hline
    \multicolumn{5}{c}{\emph{TESS light curve parameters}} \\
    $f_{0}$ [ppt] & $\mathcal{U}(-\inf,\inf)$ & \multicolumn{3}{c}{$0.033\pm 0.009$} \\
    Photometric jitter, $\ln{s^2}$ [ppt$^2$] & $\mathcal{N}(\ln{\text{var}(f_{\text{TESS}})},1)$ & \multicolumn{3}{c}{$1.21\pm 0.01$} \\
    \tess{} limb darkening coefficient, $u_1$ & $\mathcal{U}(-\inf,\inf)$ & \multicolumn{3}{c}{$0.28^{+0.33}_{-0.20}$} \\
    \tess{} limb darkening coefficient, $u_2$ & $\mathcal{U}(-\inf,\inf)$ & \multicolumn{3}{c}{$0.16^{+0.37}_{-0.28}$} \\
    $\ln{\sigma}$ [ppt] & $\mathcal{N}(\ln{\text{median}(\sigma_{f_{\text{TESS}}})},1)$ & \multicolumn{3}{c}{$-2.09^{+0.33}_{-0.36}$} \\
    \smallskip
    $\ln{\rho}$ [days] & $\mathcal{N}(\ln{50},1)$ & \multicolumn{3}{c}{$1.13^{+0.45}_{-0.44}$} \\
    \multicolumn{5}{c}{\emph{RV parameters}} \\
    $\ln{\Sigma}$ [m/s] & $\mathcal{U}(-3,3)$ & \multicolumn{3}{c}{$-0.64^{+0.74}_{-1.51}$} \\
    $P_{\text{rot}}$ [days] & $p(P_{\text{rot}}\vert H\alpha)^*$ & \multicolumn{3}{c}{$44.24^{+0.99}_{-1.00}$} \\
    $\ln{Q_0}$ & $p(\ln{Q_0}\vert H\alpha)^*$ & \multicolumn{3}{c}{$0.61^{+0.56}_{-0.55}$} \\
    $\ln{dQ}$ & $p(\ln{dQ}\vert H\alpha)^*$ & \multicolumn{3}{c}{$-0.17^{+1.81}_{-1.87}$} \\
    $\ln{f}$ & $p(\ln{f}\vert H\alpha)^*$ & \multicolumn{3}{c}{$-0.18^{+0.15}_{-0.22}$} \\
    Log jitter, $\ln{s}_{\rm RV}$ [m/s] & $\mathcal{U}(-3,3)$ & \multicolumn{3}{c}{$0.57^{+0.13}_{-0.16}$} \\
    Jitter, $s_{\rm RV}$ [m/s] & - & \multicolumn{3}{c}{$1.77^{+0.26}_{-0.26}$} \\
    \smallskip
    Systemic velocity, $\gamma_{\rm RV}$ [m/s] & $\mathcal{U}(-41650,-41630)$ & \multicolumn{3}{c}{$-41639.9424453^{+0.25}_{-0.24}$} \\    
    \multicolumn{5}{c}{\emph{Measured planetary parameters}} \\
    \hline
    && \emph{TOI-1266 b} & \emph{TOI-1266 c} & \emph{TOI-1266 d (candidate)} \\
    \hline
    Orbital period, $P$ [days] & $\mathcal{U}(-\inf,\inf)$ & $10.894841\pm 0.000011$ & $18.801611\pm 0.000053$ & $32.340\pm 0.099$ \\
    Period ratio, $P_{c}/P_b$ &-& \multicolumn{2}{c}{$1.725738\pm 0.000004$} & \\
    Period ratio, $P_{d}/P_c$ &-&& \multicolumn{2}{c}{$1.7201\pm 0.0052$} \\
    Time of mid-transit, $T_0$ [BJD - 2,457,000] & $\mathcal{U}(-\inf,\inf)$ & $2660.64605\pm 0.00055$ & $2648.84710\pm 0.00095$ & $2671.1\pm 1.0$ \\
    Transit duration, $D$ [hrs] &-& $2.071^{+0.039}_{-0.032}$ & $1.951^{+0.081}_{-0.074}$ & - \\
    Transit depth, $Z$ [ppt] &-& $3.09^{+0.14}_{-0.14}$ & $2.05^{+0.18}_{-0.17}$ & - \\
    Scaled semimajor axis, $a/R_\star$ & $p(a/R_\star\vert P,M_{\star}, R_{\star})^\dagger$ & $36.21^{+1.26}_{-1.21}$ & $52.10^{+1.82}_{-1.74}$ & $73.98^{+2.55}_{-2.37}$ \\
    Log planet-to-star radius ratio, $\ln{R_p/R_\star}$ & $\mathcal{U}(-5,0)$ & $-2.890^{+0.023}_{-0.023}$ & $-3.094^{+0.045}_{-0.043}$ & - \\
    Planet-to-star radius ratio, $R_p/R_\star$ & - & $0.0556^{+0.0013}_{-0.0013}$ & $0.0453^{+0.0020}_{-0.0019}$ \\
    Impact parameter, $b$ & $\mathcal{U}(0,1+R_p/R_\star)$ & $0.549^{+0.047}_{-0.057}$ & $0.769^{+0.030}_{-0.035}$ & \\
    $\sqrt{e}\cos{\omega}$ & $\mathcal{U}(-1,1)$ & $-0.14^{+0.17}_{-0.13}$ & $-0.04^{+0.16}_{-0.17}$ & $0.01\pm 0.15$ \\
    $\sqrt{e}\sin{\omega}$ & $\mathcal{U}(-1,1)$ & $0.05^{+0.15}_{-0.16}$ & $0.06^{+0.20}_{-0.17}$ & $0.02\pm 0.15$ \\
    Log RV semiamplitude, $\ln{K}$ [m/s] & $\mathcal{U}(-3,3)$ & $0.78^{+0.15}_{-0.18}$ & $0.21^{+0.24}_{-0.32}$ & $0.49^{+0.19}_{-0.23}$ \\
    RV semiamplitude, $K$ [m/s] & -& $2.17^{+0.34}_{-0.35}$ & $1.23^{+0.34}_{-0.34}$ & $1.63^{+0.33}_{-0.33}$ \\
    \multicolumn{5}{c}{\emph{Derived planetary parameters}} \\
    \hline
    Inclination, $i$ [deg] &-& $89.13^{+0.11}_{-0.10}$ & $89.15^{+0.06}_{-0.06}$ & - \\
    Eccentricity, $e$ &-& $<0.22^\ddagger$ & $<0.34^\ddagger$ & $<0.12^\ddagger$ \\
    Planet radius, $R_{p}$ [R$_{\oplus}$] &-& $2.62\pm 0.11$ & $2.13\pm 0.12$ & - \\
    Planet mass, $M_{p}$ [M$_{\oplus}$] & - & $4.23\pm 0.69$ & $2.88\pm 0.80$ & $4.59^{+0.96}_{-0.94}$$^{\S}$ \\ 
    Bulk density, $\rho$ [g/cm$^3$] &-& $1.3\pm 0.3$ & $1.6^{+0.6}_{-0.5}$ & - \\
    Surface gravity, $g$ [m/s$^2$] &-& $6.04\pm 1.1$ & $6.2^{+2.0}_{-1.9}$ & - \\
    Escape velocity, $v_{\text{esc}}$ [km/s] &-& $14.2^{+1.2}_{-1.2}$ & $13.0^{+1.7}_{-2.0}$ & - \\
    Semimajor axis, $a$ [au] &-& $0.0728^{+0.0011}_{-0.0011}$ & $0.1047^{+0.0016}_{-0.0016}$ & $0.1500^{+0.0023}_{-0.0024}$ \\
    Instellation flux, $F$ [$\mathrm{F}_{\oplus}$] &-& $5.5^{+1.1}_{-1.0}$ & $2.7^{+0.5}_{-0.4}$ & $1.3^{+0.3}_{-0.2}$ \\
    Equilibrium temperature$^\P$, T$_{\text{eq}}$ [K] &-& $425^{+20}_{-19}$ & $354^{+16}_{-16}$ & $297^{+14}_{-13}$ \\
    TSM$^\triangle$, & - & $138^{+35}_{-26}$ & $91^{+41}_{-25}$ & - \\
    \hline
    \multicolumn{5}{l}{$^*$ Marginalized posterior distribution from training on our dLW time series.} \\
    \multicolumn{5}{l}{$^\dagger$ $a/R_\star$ priors are based on the planet's period, and the stellar mass and radius.} \\
    \multicolumn{5}{l}{$^\ddagger$ 95\% upper limit.} \\
    \multicolumn{5}{l}{$^\S$ For the candidate planet TOI-1266 d, we only measure the minimum planet mass $M_{p,d}\sin{i}$ rather than $M_{p,d}$.} \\
    \multicolumn{5}{l}{$^{\P}$ Equilibrium temperature calculations assume zero albedo and perfect heat redistribution.} \\
    \multicolumn{5}{l}{$^{\triangle}$ Transmission Spectroscopy Metric \citep{Kempton_2018}.} \\
\end{tabular}
\end{table*}

\subsection{Evidence for a Third Planet Around TOI-1266} \label{sect:evidence}
In Section~\ref{sect:analysis:global} we considered four RV+transit models with either two or three planets, which we denote as 2p and 3p, respectively. The third planetary component in our 3p models is used to model the 32.3-day signal seen in the GLS of our RV time series. We also considered two versions of our 2p and 3p models that do and do not include a stellar activity component in the form of a trained GP. For each model $M$, we estimate its Bayesian model evidence $\mathcal{Z}_M$ using the Perrakis estimator, for which we use the model's joint posterior from our MCMC calculations as an importance sampler \citep{Perrakis_2014}. We then compute Bayes factors, or evidence ratios, to determine the favoured model given our RV data. Our results are reported in Table~\ref{tab:lnZ}.

We find that in both our 2p and 3p models, the inclusion of a GP is favoured.\footnote{We note that given the difficulty in accurately calculating $\mathcal{Z}_M$ values, we follow the recommendation from \cite{Nelson_2020} that increasingly complex models should have Bayes factors $>10^3$ in order to be robustly favoured over comparatively simplistic models.} Despite TOI-1266 being an inactive star, the inclusion of a GP activity component produces a more complete model. We also find that our 3p models are heavily favoured over our 2p models, regardless of whether a GP is included. Specifically, $\mathcal{Z}_{3p+\mathrm{GP}} / \mathcal{Z}_{2p+\mathrm{GP}} = 10^{6.4}$ and $\mathcal{Z}_{3p} / \mathcal{Z}_{2p} = 10^{5.6}$. This suggests that the 32.3-day signal in the GLS of the RVs is well-described by a Keplerian orbital solution. We consider this to be tentative evidence for a third planet candidate in the system whose planetary parameters are included in Table~\ref{tab:results}.

\subsection{TTV analysis}
We do not find strong evidence for TTVs in the TESS photometry. This result is wholly consistent with our measured RV masses given the relatively large photometric rms of the individual TESS transit events, which inhibit the precise measurement of individual transit times. This result is consistent with previous findings based on TESS PM data \citepalias{Stefansson_2020}. However, with the planets’ masses constrained by our RV measurements, here we supplement our pseudo-Keplerian analysis by modeling the lack of TESS TTVs to place stringent upper limits on the transiting planets’ eccentricities following the formalism of \cite{Hadden_2019}. 

Since the transit times of TOI-1266 c may be influenced by dynamical interactions with
the candidate planet d, we focus on deriving eccentricity constraints by modeling TOI-1266 b’s transit times only. We do so by adopting a linear model approach as described in \cite{Linial_2018}. We are sensitive to the magnitude of `combined eccentricity' in this model, $|\mathcal{Z}_{b,c}|\approx 1/\sqrt{2}\sqrt{e_b^2 + e_c^2-2e_be_c\cos(\varpi_b-\varpi_c)}$, which depends on the planet eccentricities $e_i$ and arguments of periastron $\varpi_i$. We use the \texttt{emcee} package \citep{foremanmackey13} to sample the likelihood of TOI-1266 b's transit times under this model, which is conditioned on the RV masses reported in Table~\ref{tab:results}. We conclude that $|\mathcal{Z}_{b,c}| < 0.08$ (0.11) at 95\% (99\%) confidence, which is more stringent than our eccentricity constraints derived from the RVs alone. 

\subsection{Transit Depth Discrepancy} \label{sect:analysis:transit}
As introduced in Section~\ref{sect:obs:tess} and illustrated in Figure~\ref{fig:transits}, our initial inspection of the PDCSAP transit light curves revealed transit depth discrepancies for both transiting planets between TESS's PM and EM. Phase-folding the PM and EM light curves based on the linear ephemerides from our preliminary analysis revealed that TOI-1266 b's transit depth increases by $1.9\sigma$ from $2.86^{+0.12}_{-0.14}$ ppt to $3.09^{+0.14}_{-0.14}$ ppt. More significantly, the transit depth of TOI-1266 c increases by $3.5\sigma$ from $1.45^{+0.18}_{-0.17}$ ppt to $2.05^{+0.18}_{-0.17}$ ppt. We checked that these distinct transit depths persist across multiple transit events in the PM and EM and are not dominated by any single event that may result from a transient event (e.g. a solar system object occultation).

We seek to resolve these discrepancies by considering five possible explanations: variable flux dilution, orbital precession, residual artifacts in the PDCSAP light curve, evolution in stellar activity, and stochastic effects between the PM and EM TESS data. In the following subsections we rule out the first three scenarios and ultimately attribute the observed transit depth discrepancy to a marginal increase in stellar activity during the TESS EM and to stochastic effects in the TESS PM data that limit the accuracy of recovered low S/N signals with the ephemeris of TOI-1266 c.

\subsubsection{Variable Flux Dilution?} \label{sect:dilution}
As discussed in Section~\ref{sect:obs:tess}, we accounted for dilution in our TPF and FFI-extracted light curves using the known positions and TESS magnitudes of neighbouring sources from the TICv8 \citep{Stassun_2019}. Yet variable dilution between the PM and EM can be immediately ruled out as the culprit for the transit depth discrepancies because its impact on both planets would be equal. The transit depth ratios between the PM and EM are $0.88\pm 0.02$ and $0.71\pm 0.04$ for TOI-1266 b and c, respectively. From these differing levels of depth increases, we conclude that variable flux dilution is not responsible for the transit depth discrepancies.

\begin{table*}
  \centering
  \caption{Comparative transit parameters across different observing facilities and epochs. \label{tab:depths}}
  \begin{tabular}{lccccccc}
    \hline
    Facility & Bandpass & Photometric & $T_0$ & Transit & Transit & Impact & Ref. \\
    && rms [ppt] & [BJD - 2,457,000] & depth [ppt] & duration [hr] & parameter & \\
    \hline
    \multicolumn{8}{c}{\it{TOI-1266 b}} \\
    \smallskip
    TESS PDCSAP PM & $T$ & 1.843 & $1691.0051\pm 0.0008$ & $2.86^{+0.12}_{-0.14}$ & $2.138^{+0.043}_{-0.034}$ & $0.511^{+0.057}_{-0.078}$ & 1 \\ 
    \smallskip
    TESS TPF PM & $T$ & 2.001 & $1691.0051\pm 0.0007$ & $2.73^{+0.11}_{-0.12}$ & $2.121^{+0.036}_{-0.033} $ & $0.522^{+0.052}_{-0.064} $ & 1 \\ 
    TESS FFI PM & $T$ & 0.475 & $1691.0064\pm0.0009$ & $2.70^{+0.27}_{-0.29}$ & $2.10^{+0.11}_{-0.11} $ & $0.54^{+0.12}_{-0.20} $ & 1 \\ 
    \hline
    \smallskip
    TESS PDCSAP EM & $T$ & 1.827 & $2420.9624\pm 0.0007$ & $3.12^{+0.14}_{-0.14}$ & $2.068^{+0.039}_{-0.032}$ & $0.566^{+0.047}_{-0.057}$ & 1\\  
    \smallskip
    TESS TPF EM & $T$ &2.002  & $2420.9627\pm 0.0007$ & $3.11^{+0.13}_{-0.14}$ & $2.050^{+0.041}_{-0.036} $ & $0.578^{+0.046}_{-0.057} $ & 1 \\ 
    \smallskip
    TESS FFI EM & $T$ & 0.832 & $2420.9627\pm 0.0009$ & $3.14^{+0.22}_{-0.23}$ & $2.089^{+0.076}_{-0.067} $ & $0.552^{+0.067}_{-0.096} $ & 1 \\ 
    \smallskip
    Palomar/WIRC & $J$ & 2.351 & $-^*$ & $3.24^{+0.51}_{-0.53}$ & $2.01^{+0.18}_{-0.18}$ & $0.520^{+0.088}_{-0.088}$ & 1 \\
    \hline
    \multicolumn{8}{c}{\it{TOI-1266 c}} \\
    \smallskip
    TESS PDCSAP PM & $T$ & 1.843 & $1689.9582\pm 0.0010$ & $1.30^{+0.18}_{-0.17}$ & $2.33^{+0.18}_{-0.27}$ & $0.61^{+0.12}_{-0.13}$ & 1\\ 
    \smallskip
    TESS TPF PM & $T$ & 2.001 & $1689.9603\pm 0.0009$ & $1.58^{+0.15}_{-0.14}$ & $1.891^{+0.079}_{-0.063}$ & $0.787^{+0.025}_{-0.033} $ & 1 \\ 
    \smallskip
    TESS FFI PM & $T$ & 0.475 & $1689.9609\pm 0.0010$ & $1.48^{+0.44}_{-0.35}$ & $1.94^{+0.42}_{-0.29} $ & $0.77^{+0.10}_{-0.23} $ & 1 \\ 
    APO/ARCTIC & $i'$ & 0.643 & $1689.962\pm 0.005$ & $1.37^{+0.44}_{-0.46}$ & $1.98^{+0.10}_{-0.19} $& $0.730^{+0.033}_{-0.070}$ & 2 \\
    \hline
    \smallskip
    TESS PDCSAP EM & $T$ & 1.827 & $2423.2223\pm 0.0009$ & $2.07^{+0.18}_{-0.17}$ & $1.949^{+0.081}_{-0.074}$ & $0.776^{+0.030}_{-0.035}$ & 1 \\  
    \smallskip
    TESS TPF EM & $T$ & 2.002 & $2423.2237\pm 0.0009$ & $1.94^{+0.18}_{-0.17}$ & $1.937^{+0.086}_{-0.079} $ & $0.778^{+0.030}_{-0.036} $ & 1 \\ 
    \smallskip
    TESS FFI EM & $T$ & 0.832 & $2423.2212\pm 0.0010$ & $2.06^{+0.34}_{-0.28}$ & $1.94^{+0.13}_{-0.13} $ & $0.780^{+0.047}_{-0.055} $ & 1 \\ 
    \smallskip
    Palomar/WIRC & $J$ & 1.493 & $-^*$ & $1.95^{+0.18}_{-0.20}$ & $1.94^{+0.17}_{-0.16}$ & $0.702^{+0.049}_{-0.055}$ & 1 \\
    \hline
    \multicolumn{8}{l}{\textbf{References:} 1) this work 2) \citetalias{Stefansson_2020}.} \\
    \multicolumn{8}{l}{$^*$ WIRC transit times are excluded here and are reserved for a forthcoming detailed TTV analysis (Greklek-McKeon et al. in prep.)} \\
  \end{tabular}
\end{table*}

\subsubsection{Orbital Precession?}
Gravitational interactions between planets in compact systems can cause precession of the planets' orbital planes, which can produce an observable signature by altering a transiting planet's impact parameter \citep{Miralda_2002}. Changing the impact parameter can result in the occultation of differentially limb-darkened transit chords, which may be able to explain the transit depth discrepancy. This effect has been observed in the K2-146 system containing two sub-Neptunes in a 3:2 resonance \citep{Hamann_2019}. However in comparison, the TOI-1266 planets are near a higher order period ratio (i.e. 5:3) that produces weaker gravitational interactions such that we do not expect to see significant orbital precession. We test this by revisiting our transit light curve fits from the PM and EM (see Section~\ref{sect:analysis:global}) and noting that each planet's transit duration are consistent within $1\sigma$ between the PM and EM. We conclude that the TOI-1266 system does not show evidence for orbital precession over the 16-month interval between the PM and the EM such that orbital precession cannot explain the observed transit depth discrepancy.

\begin{figure}
    \centering
    \includegraphics[width=0.96\hsize]{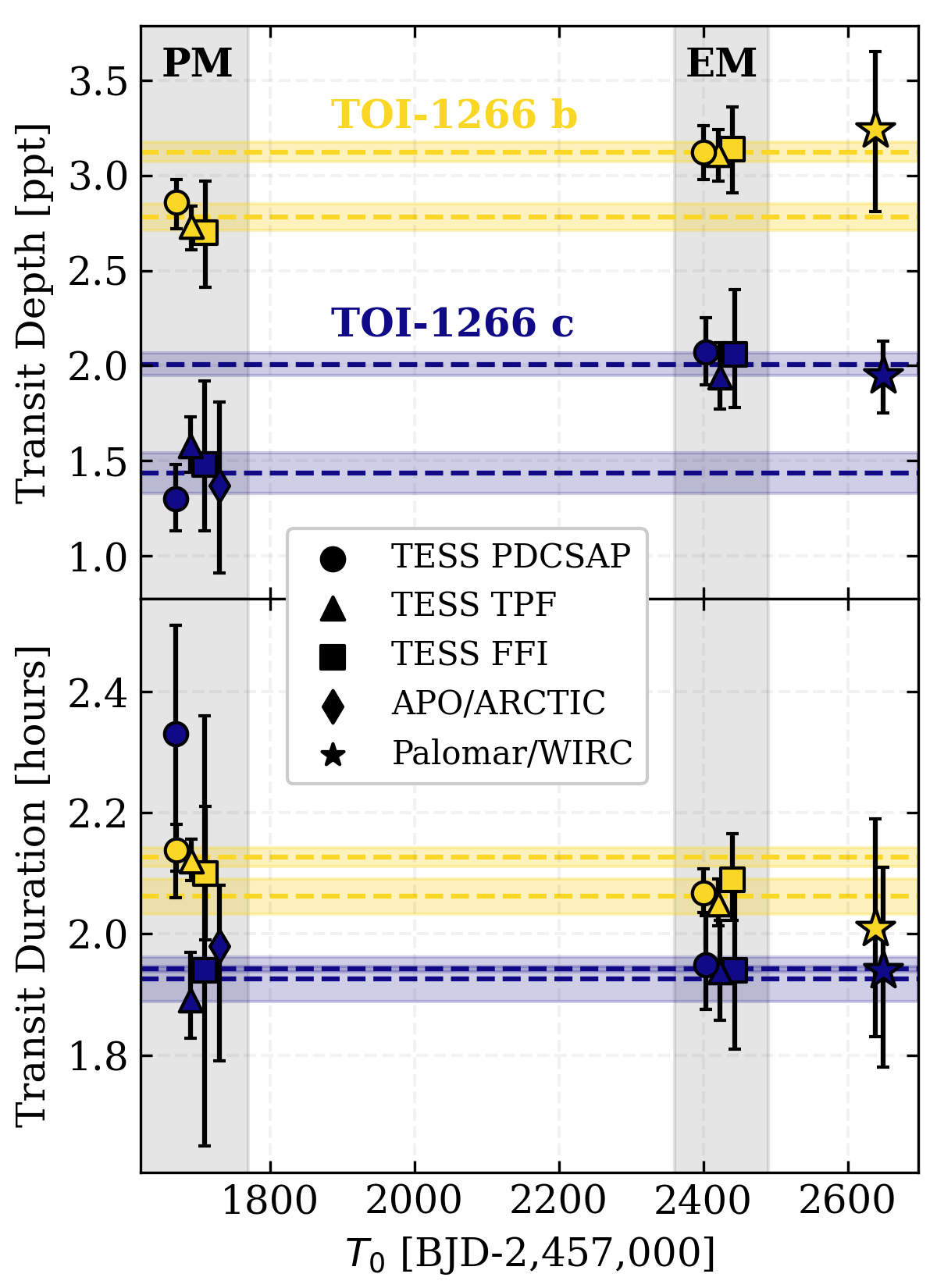}
    \caption{Observed transit depths and durations for TOI-1266 b (yellow) and c (blue) as a function of time and based on different observing facilities or different TESS data reductions (different markers). The horizontal bands depict the median and standard deviations of each planet's depth and duration values across all facilities but in the PM and EM epochs separately. One exception is the transit duration of TOI-1266 c from the TESS PM PDCSAP light curve, which we omit due to its anomalous nature when compared to the other extraction methods and facilities.}
    \label{fig:depths}
\end{figure}

\subsubsection{Residual Artifacts in the PDCSAP Light Curve?} \label{sect:altdata}
We investigate whether the transit depth discrepancies are unique to the SPOC's PDCSAP light curve construction by considering alternative TESS data reductions and precise transit light curves that were not obtained with TESS. As outlined in Section~\ref{sect:obs:tess}, we used our custom light curve extractions from each TESS sector based on the TPFs and FFIs. We corrected each light curve for dilution and detrended using a GP following our methodology in Section~\ref{sect:analysis:global}. 

Figure~\ref{fig:depths} and Table~\ref{tab:depths} report each planet's transit depth and duration measured by TESS, or from the ground, in the TESS PM, EM, and post-EM. Across the three TESS extractions considered (i.e. PDCSAP, TPF, and FFI), both planet's transit depths increase from the PM to the EM epochs while their transit durations remain constant. These findings are consistent across the different TESS extractions. 

To verify that these depth discrepancies are not unique to TESS, we sought to obtain high-precision transit observations from the ground using diffuser-assisted imagers. While no such observations are known to exist for TOI-1266 b around the PM epochs, we retrieved the full transit observation of TOI-1266 c from \citetalias{Stefansson_2020} taken with the ARCTIC camera (Section~\ref{sect:arctic}). These data confirm the PM transit depth and duration of TOI-1266 c observed by TESS (Figure~\ref{fig:depths}). We followed-up with additional transit observations of TOI-1266 b and c in the $J$-band from the WIRC camera (Section~\ref{sect:wirc}). These ground-based data are consistent with the deeper transits seen by TESS in the EM versus the PM, and confirm the invariability of both planets' transit durations over time. We conclude that the observed depth discrepancies are not unique to TESS and that they are robust to the data reduction steps and across different observing facilities.

\begin{figure}
    \centering
    \includegraphics[width=0.99\hsize]{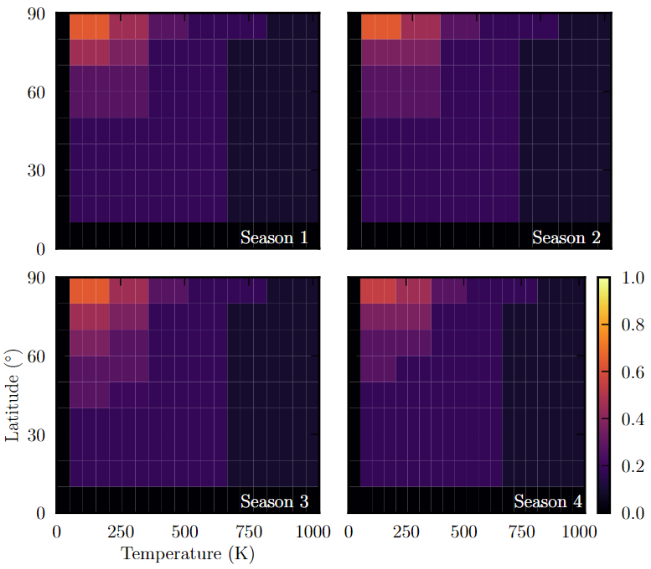}
    \caption{The results of our active region simulations using \texttt{SOAP 2.0}. The panels depict the fraction of plage configurations that are consistent with our observations in each of our spectroscopic follow-up observing seasons and as a function of latitudinal position and temperature contrast with the photosphere (i.e. marginalized over longitude and plage size). Cool plages at high latitudes are allowed in each observing season with season 3 exhibiting the highest number of allowed models, indicative of a slight increase in TOI-1266's magnetic activity level.}
    \label{fig:soap}
\end{figure}

\subsubsection{Evolution in Stellar Activity?} \label{sect:analysis:soap}
We investigate whether active region (AR) evolution could plausibly explain the observed transit depth discrepancies. The TESS photometric rms does not vary significantly between the PM and EM (Table~\ref{tab:depths}), such that we can rule out dark starspots as the culprit. Here we consider bright plages by simulating their impact on the TESS photometry, HARPS-N LBL RVs (less the known planets), CCF FWHM, and CCF BIS timeseries using an adapted version of \texttt{SOAP 2.0} (\citealt{Dumusque_2014}; Wurmser et al. in prep.). We explored a grid of $N=43,890$ plage sizes ($0-0.65\,R_\star$), temperature contrasts ($0-1000$ K), latitudes ($0-90\deg$), and longitudes ($0-360\deg$), and determined the activity-induced rms values on each of the aforementioned time series. Planet population studies suggest that small planets are preferentially well-aligned with their host stars' rotation axes \citep{Albrecht_2022} such that we assume TOI-1266's rotation axis is perpendicular to our line-of-sight. We adopt \prot{}$=30$ days as these calculations were performed before the identification of the probable rotation period at 44.6 days. Fortunately, our preliminary tests sampled orbital periods from 30-60 days and revealed that our results are insensitive to the exact choice of \prot{.}

We compare our grid of plage models to our data in each of our four observing seasons of spectroscopic follow-up (see Figure~\ref{fig:rvs}). In this way, we are sensitive to changes in the plage properties that may produce the observed transit depth discrepancies. In each model, we consider plages to be observationally ``allowed'' if each of the model rms values in photometry, RVs, FWHMs, and BISs are less than the observed rms of the corresponding time series. Figure~\ref{fig:soap} shows the results of this exercise. We find that small, cool plages at high latitudes ($R_{\mathrm{plage}}/R_\star \lesssim 0.17$, $\Delta T \lesssim 300$ K, $\phi \gtrsim 80$\degree) are consistent with the data in all four observing seasons. However, a larger fraction of models is consistent with the data in season 3 wherein plages, of a fixed size, extend down to lower latitudes in season 3 (UT 2021 Dec 18 to 2022 July 14). We conclude that season 3 exhibits the highest level of activity out of the four seasons. Season 3 corresponds to epochs that overlap with sectors 48 and 49 of the TESS EM wherein we witness the transit depth increases compared to the PM. 

Our finding that plages on TOI-1266 are localized at high latitudes is consistent with the findings of \cite{Morales_2010} who analyzed a set of eclipsing binaries and concluded that polar magnetic ARs on low mass stars are a preferred solution to explain differences between stellar model predictions and EB measurements of fundamental stellar parameters. The existence of polar plages on TOI-1266 also provides a plausible explanation for the more severe transit depth discrepancy exhibited by TOI-1266 c than b due to its larger impact parameter (i.e. $b_c=0.77$; $b_b=0.55$).

Because TOI-1266's magnetic activity is dominated by bright plages rather than by dark spots, the impact of ARs on the observed transit depths will be to increase the transit depth relative to an unspotted photosphere. \cite{Morris_2018} suggest a framework to disentangle the effects of $R_p/R_\star$ and unocculted ARs to recover robust planetary radii from transit depth measurements. The basis of the framework is that while the true value of $R_p/R_\star$ affects both the transit depth and the ingress/egress durations, only the depth is impacted by unocculted ARs. The ingress/egress durations may therefore be used to infer accurate $R_p/R_\star$ values if the transit light curve has sufficient S/N and limb-darkening is well-constrained. In light of the possibility of elevated plage coverage in observing season 3, we attempted to use this framework to measure each planet's $R_p/R_\star$ value in the full TESS EM, the individual EM sectors, and in the diffuser-assisted WIRC transits.
We find that in all light curves, the photometric S/N values are too low to precisely recover the ingress/egress times such that we measure planetary radii that are consistent between spotted and unspotted photospheres. That is, while there exists some evidence that TOI-1266's activity level increases slightly toward the TESS EM epochs, this does not have a demonstrable impact on the observed transit depths.

\subsubsection{Stochastic Effects Between the PM and EM TESS Data} \label{sect:injrec}
We consider whether there are stochastic effects in either the PDCSAP PM or EM data that drive the disagreement in transit depths between their respective light curves. Here we focus on TOI-1266 c, which shows a greater depth discrepancy than b. We proceed by injecting transit signals into our cleaned TESS data, recovering them within our model fitting framework (Section~\ref{sect:analysis:global}), and comparing the recovered transit depths to those injected. We sample transit depths from a log uniform grid spanning $Z_{\mathrm{inj}} \in [0.5,3]$ ppt. We separately consider two sets of ephemerides for our injected planets: i) randomized mid-transit times and orbital periods between 15-25 days and ii) injected planets with the same ephemeris as TOI-1266 c. We inject these signals into the cleaned PDCSAP light curves after removing our MAP GP and transit models from the PM and EM, respectively.

\begin{figure}
    \centering
    \includegraphics[width=0.9\hsize]{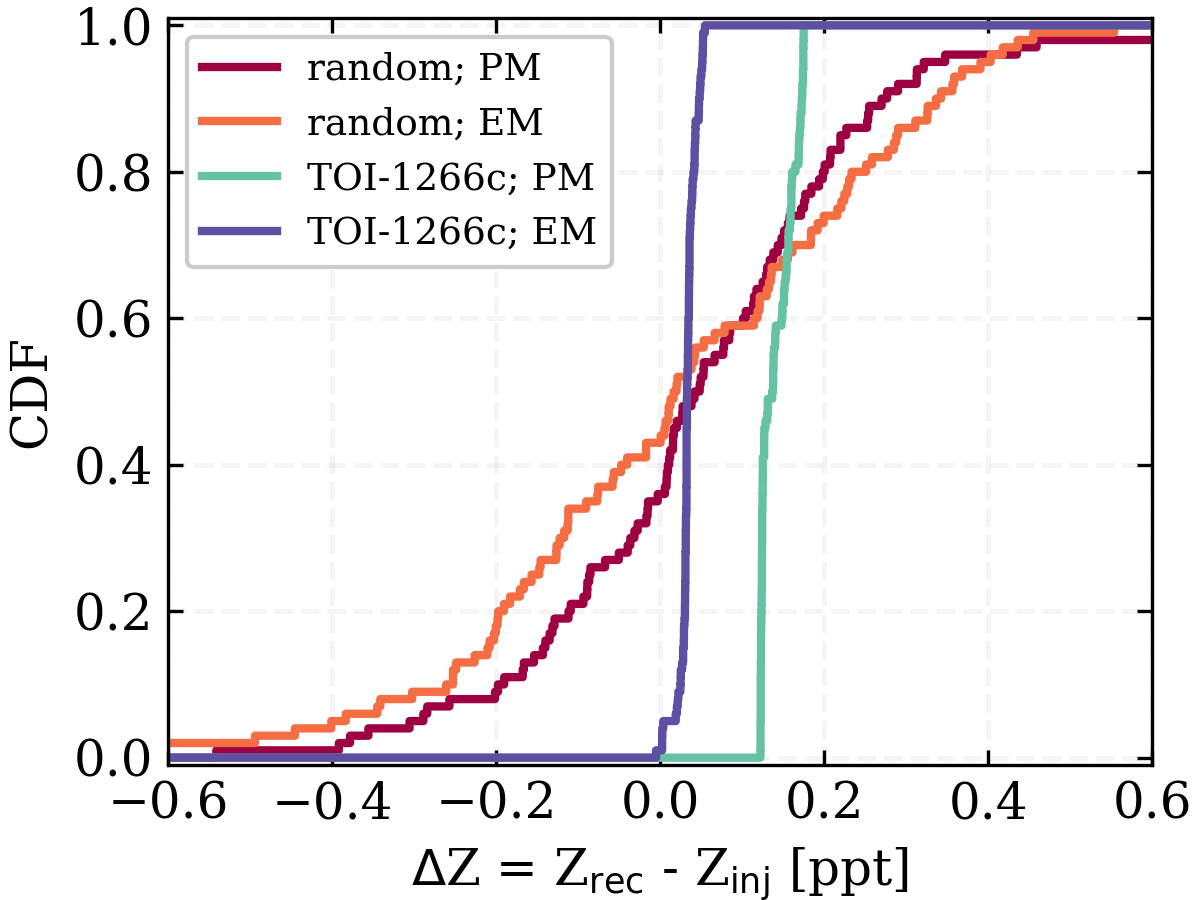}
    \caption{Cumulative distributions of transit depth measurement accuracies $\Delta Z$ from a set of injection/recovery tests. Four sets of models are shown here comparing results from the TESS PM and EM, and for two different sets of injected planet ephemerides: i) with random orbital periods and mid-transit times and ii) with the linear ephemeris of TOI-1266 c.}
    \label{fig:deltaZ}
\end{figure}

Figure~\ref{fig:deltaZ} compares the resulting distributions of transit depth differences, $\Delta Z = Z_{\mathrm{rec}}-Z_{\mathrm{inj}}$, for the PM and EM and for each set of injected ephemerides. We treat $\Delta Z$ as a measure of the accuracy with which transit depths are recovered. We find that when planets are injected with random ephemerides, the $\Delta Z$ distributions are Gaussian-distributed in both the PM and EM, and with similar rms values of 0.21 and 0.25 ppt, respectively. The values are comparable to the photometric rms in the PM and EM light curves (Table~\ref{tab:depths}). This implies that transiting planets with random transit times and orbital periods can be accurately recovered in both the PM and EM PDCSAP light curve. Conversely, we find that injected planets with the same ephemeris as TOI-1266 c exhibit depth measurement accuracies that differ between the PM and EM. This effect is not due to the smaller number of transits observed because our randomized injected ephermides have comparable periods to TOI-1266 c. We note that the depth of a c-like planet is more accurately recovered in the EM than in the PM. This is evidenced by the distribution of $\Delta Z$ in the EM being narrowly distributed around 0.03 ppt, whereas the typical accuracy of transit depths measured in the PM is considerably larger at 0.14 ppt.

Recall that the observed transit depth discrepancy for TOI-1266 c is $0.6$ ppt, which is greater than the level of inaccuracy that our injection/recovery tests suggest in the PM. However our results do indicate that the measured transit depth of planet c is more reliable in the EM than in the PM. Our results also indicate that this is a stochastic effect related to the exact ephemeris of TOI-1266 c because signals injected with random ephemerides do not show the same systematic overestimate in transit depth. This may explain why the transit depth discrepancy is significant for planet c ($3.5\sigma$) while it is marginal for b ($1.9\sigma$).

\subsubsection{Conclusions Regarding the Transit Depth Discrepancy} \label{sect:analysis:conclude}
We conclude that variable flux dilution, orbital precession, and residual artifacts in the PDCSAP light curve cannot explain the observed transit depth discrepancies between the TESS PM and EM. While there is some evidence that the star's magnetic activity level increased during our third observing season of spectroscopic follow-up, which aligns with the EM epochs that show increased depths, the impact is small and is likely an incomplete explanation of the apparent transit depth increase. There is also evidence that the transit depth of a planet with TOI-1266 c's ephemeris is recovered more accurately in the EM versus the PM. Although the level of improvement is similarly insufficient to explain the observed transit depth discrepancy. We conclude that part of the transit depth discrepancy can be explained by a small increase in the stellar activity and to stochastic effects that limit the accuracy with which TOI-1266 c's transit depth is measured in the PM. We proceed with adopting our fit results from the EM but caution that there is still room for further clarification on the transit depth of TOI-1266 c, requiring additional follow-up with ultra-precise photometers.

\subsection{Transiting Planet Search} \label{sect:tls}
We conducted an independent search for transiting planets using the Transit Least Squares algorithm \citep[\texttt{TLS};][]{Hippke_2019}. We searched for transit-like signals between 19-50 days in the combined PDCSAP light curve after detrending for residual systematics and the two known transiting planets. We chose our period bounds because we do not expect to have enhanced sensitivity to transiting planets with $P<P_c=18.8$ days compared to the TESS team, and because the transit probability drops below 1\% at $\sim 50$ days. TOI-1266 b and c are remarkably coplanar ($\Delta i < 0.1^\circ$) and a third coplanar planet would no longer exhibit a transiting configuration beyond 27.1 days. We also conducted a focused TLS search between 31.9-32.7 days to check if the RV planet candidate revealed in Section~\ref{sect:analysis:gls} is transiting. We also conducted a by-eye search for this planet candidate based on its assumed linear ephemeris from our RV analysis.

Our search yielded no new transit-like signals. We proceeded with injecting synthetic planetary signals into our TESS data to quantify our sensitivity to transiting planets. We sampled orbital periods logarithmically between 19-50 days, uniform times of mid-transit, planet radii logarithmically between $0.5-3\, \mathrm{R}_{\oplus}$, and impact parameters linearly between $0-1$. We constructed $10^4$ synthetic light curve realizations using the \texttt{batman} package \citep{kreidberg15} to compute the transit models. We then processed each light curve using the \texttt{TLS} and recovered planets with Signal Detection Efficiencies $>6$ \citep{Hippke_2019} within 5\% of the injected period. 

\begin{figure}
    \centering
    \includegraphics[width=\hsize]{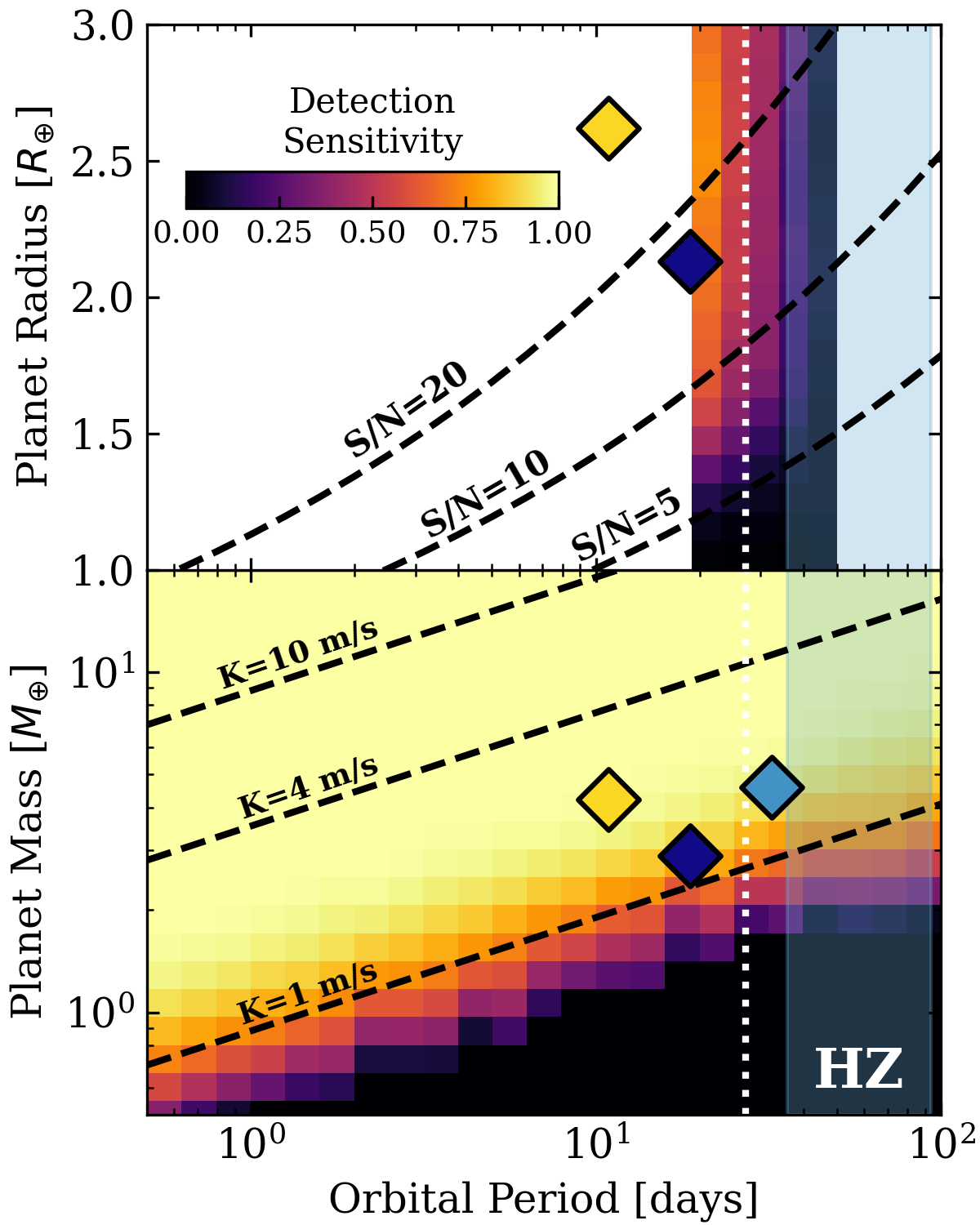}
    \caption{Our detection sensitivity to additional planets in the TOI-1266 system based on our PDCSAP TESS photometry (top panel) and HARPS-N RVs (bottom panel). We only search for transiting planets beyond the orbit of TOI-1266 c ($P_c=18.8$ days) and out to 50 days where our sensitivity becomes consistent with 0\%. The dashed lines depict lines of constant transit S/N (upper panel) and constant RV semiamplitude (lower panel). The diamond markers highlight TOI-1266 b (yellow), c (dark blue), and the non-transiting planet candidate d (light blue). The vertical dotted line depicts the maximum orbital period that a hypothetical third planet could have and still exhibit a transiting configuration if it was coplanar with TOI-1266 b and c. The shaded region highlights the habitable zone between 36 and 95 days.}
    \label{fig:sens}
\end{figure}

The top panel of Figure~\ref{fig:sens} depicts our detection sensitivity to transiting planets. We do not correct for the geometric probability given the preference of compact multi-planet systems to exhibit small mutual inclinations $\lesssim 1-2^\circ$ \citep{ballard16}. We confidently rule out transiting planets down to $1.4\, \mathrm{R}_{\oplus}$ within 23 days but lack sensitivity to planets within the habitable zone (HZ) bounded by the water-loss and maximum greenhouse limits \citep[36-95 days;][]{Kopparapu_2013}.

\subsection{Radial Velocity Planet Search}

\begin{figure*}
    \centering
    \includegraphics[width=\hsize]{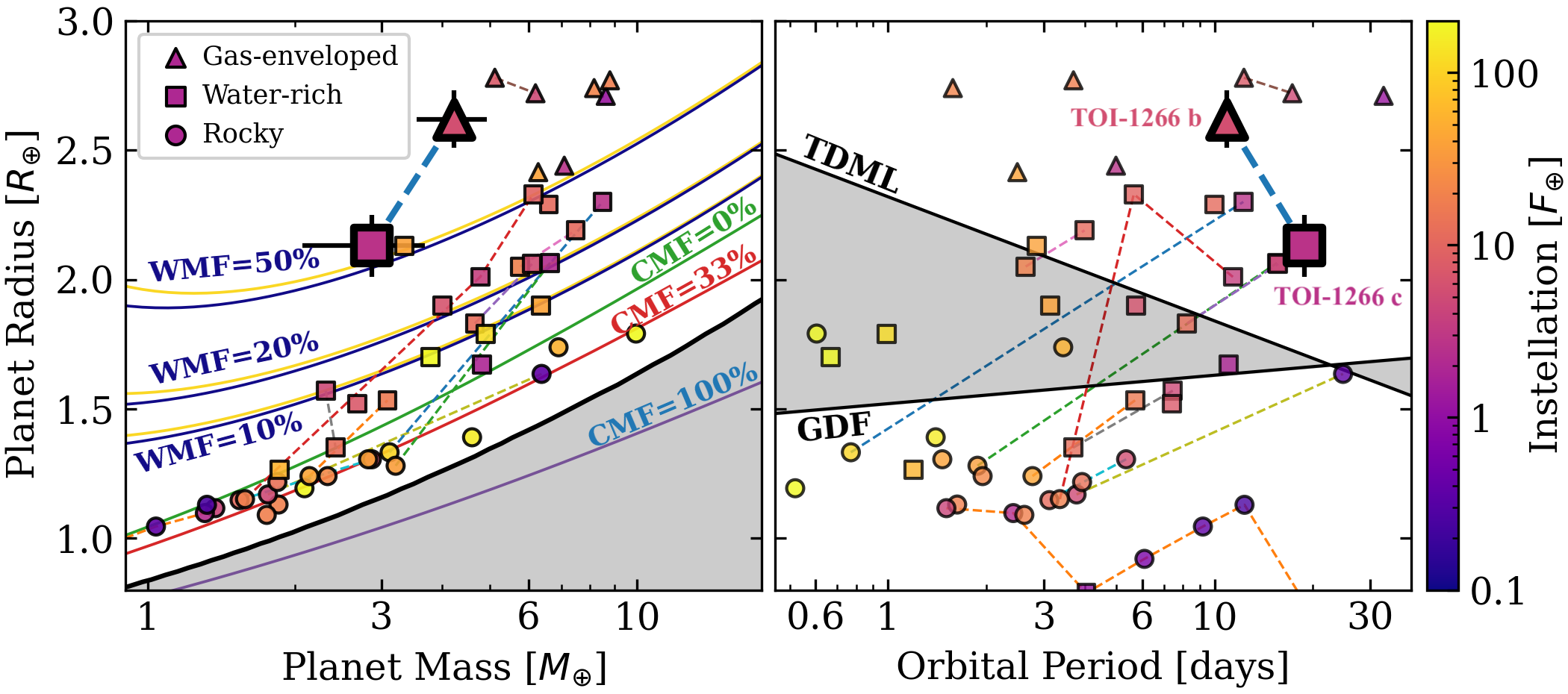}
    \caption{Left panel: the mass-radius-instellation diagram for 
    small transiting planets around M dwarfs. The marker symbols depict planets with different bulk compositions based on their masses and radii with the TOI-1266 planets, highlighted with bold markers. The solid curves illustrate interior structure models from \citealt{Zeng_2013} with iron core mass fractions (CMF) of 0\%, 33\% (i.e. Earth-like), and 100\%, and water-rich models from \citealt{Aguichine_2021} with water mass fractions (WMF) of 10\%, 20\%, and 50\% at the equilibrium temperatures of TOI-1266 b (yellow) and c (blue). The shaded region highlights the densest small planets predicted by models of maximum collisional stripping \citep{Marcus_2010}. Right panel: the same planet sample in period-radius-instellation space. The shaded region is bounded by model predictions of the location of the M dwarf radius valley from thermally driven mass loss (TDML) and gas-depleted formation (GDF) \citep{Cloutier_Menou}. Multi-planet systems are connected by the dashed lines in both panels.}
    \label{fig:mr}
\end{figure*}

We conducted a similar exercise to assess our detection sensitivity to planets in our HARPS-N RV timeseries. We sampled the injected planetary parameters identically as in Section~\ref{sect:tls} with the exception that because we do not require injected planets to be transiting, we revise our impact parameter sampling to be linearly sampled from $\pm a/R_\star$, where $a$ is determined by the sampled value of the orbital period. These sensitivity calculations therefore represent conservative estimates as we are not preferentially sampling coplanar systems. Additionally, we logarithmically sample planet masses between $0.5-20\, \mathrm{M}_\oplus$. We constructed $10^4$ synthetic RV time series by computing the Keplerian orbit for each synthetic planet and injected individual planets into our HARPS-N residuals (i.e. bottom panel in Figure~\ref{fig:rvs}). We deem successful recoveries of injected planets to be those that exhibit a FAP $\leq 0.1$\% in the GLS within 5\% of the injected period and have $\Delta$BIC $<-10$, where $\Delta$BIC is the Bayesian Information Criterion comparing the injected Keplerian model to the null hypothesis (i.e. a flat line). The bottom panel of Figure~\ref{fig:sens} depicts our detection sensitivity to planets in our RV time series as a function of orbital period and true mass. We confidently rule out planets down to $3\, \mathrm{M}_\oplus$ within 20 days and planets down to $4\, \mathrm{M}_{\oplus}$ within the HZ, noting however that we are only sensitive to the planet's minimum mass if it is only detected in the RV data.

\section{Results and Implications for the System's Formation} \label{sect:disc}
\subsection{Fundamental Planetary Parameters} \label{sect:mr}
We measure orbital parameters for the transiting planets TOI-1266 b and c that are consistent with values reported in their discovery papers (\citetalias{Demory_2020,Stefansson_2020}). Conversely, our transit analysis suggests increased planet radii that arise from increased transit depths exhibited by the TESS EM data compared to the PM. In Section~\ref{sect:analysis:transit} we deemed the PM to suffer from stochastic effects that weaken the accuracy of TOI-1266 c's measured transit depth. We consequently adopt the results from our EM model in the forthcoming discussion.

We recover planetary radii of $R_{p,b}= 2.62\pm 0.11\, \mathrm{R}_{\oplus}$ and $R_{p,c} = 2.13\pm 0.12\, \mathrm{R}_{\oplus}$. We also measure the masses of both planets using data from the HARPS-N spectrograph ($M_{p,b} =4.23\pm 0.69\, \mathrm{M}_{\oplus}$ and  $M_{p,c}=2.88\pm 0.80\, \mathrm{M}_{\oplus}$). Our results correspond to $24\sigma$ and $19\sigma$ radius measurements and $6.1\sigma$ and $3.6\sigma$ mass measurements for TOI-1266 b and c, respectively. We also uncover a third planet candidate in the system and measure its minimum planet mass $M_{p,d}\sin{i}=4.59^{+0.96}_{-0.94}\, \mathrm{M}_{\oplus}$ ($4.8\sigma$). If real, TOI-1266 d would be a temperate super-Earth or sub-Neptune located just inside the inner edge of the water-loss HZ at an instellation of $F_d = 1.3^{+0.3}_{-0.2}\, \mathrm{F}_{\oplus}$.

The left panel in Figure~\ref{fig:mr} compares the masses and radii of TOI-1266 b and c to the current population of small planets around M dwarfs with well-characterized masses (i.e. $\geq 3\sigma$). The TOI-1266 planets occupy a unique region of the mass-radius parameter space as they represent the lowest density known sub-Neptunes around M dwarfs in the size range of $1.7-3\, \mathrm{R}_{\oplus}$. If we assume that both planets have Earth-like cores with core mass fractions CMF of 33\%, and are enveloped in H$_2$ gas, then at their present-day equilibrium temperatures of 425 K and 354 K, respectively, we infer envelope mass fractions of $X_{\mathrm{env},b}=5.5\pm 0.7$\% and $X_{\mathrm{env},c}=3.5\pm 0.8$\%.

However, multiple lines of recent empirical evidence have suggested that unlike around FGK-stars, sub-Neptunes around M dwarfs are inconsistent with being gas-enveloped terrestrials \citep{Cloutier_Menou,DiamondLowe_2022,Luque_2022,Piaulet_2023,Cherubim_2023}. Instead, the small, close-in planets around M dwarfs that are too low density to be super-Earths are likely irradiated ocean worlds with substantial water mass fractions (WMF) and negligible envelope mass fractions. Adopting the water-rich mass-radius relations from \cite{Aguichine_2021}, which are state-of-the-art by their use of a multi-phase and non-isothermal equation of state for water, we find that WMF$_b>80$\% at 95\% confidence\footnote{This includes WMF$_b> 100$\% (i.e. TOI-1266 b must be gas-enveloped).} and WMF$_c=59\pm 14$\%. The mass-radius relationships from \cite{Aguichine_2021} are only computed down to equilibrium temperatures of 400 K, which required us to extrapolate to compute WMF$_{c}$ at $T_{\mathrm{eq}}=354$ K. 

We note that at solar composition, the WMF beyond the water snow line is expected to be at most 50\% \citep{Lodders_2003} such that TOI-1266 c may be a bona-fide water-world. We highlight that while we cannot definitively classify TOI-1266 c as a water-world with solely the data herein, we present atmospheric escape models in Section~\ref{sect:radval} that provide additional evidence that planet c likely has a water-rich bulk composition. However, we note that breaking this degeneracy and establishing the true bulk composition of TOI-1266 c will require detailed atmospheric characterization to measure its atmospheric mean molecular weight. We report the TSM values of both planets in Table~\ref{tab:results}.

We note that if TOI-1266 c turns out to be a confirmed water-world, the system's architecture would be one-of-a-kind as the only M dwarf system to feature a gas-enveloped sub-Neptune orbiting interior to a rocky planet or a water-world (c.f. right panel of Figure~\ref{fig:mr}). The only comparable ``inverted'' planet pairs are TOI-270 c and d, which are both consistent with being water worlds \citep{VanEylen_2021}, and Kepler-26 b and c, which are both consistent with being gas-enveloped. While the planet formation process is a stochastic one, ``inverted'' architectures with planets of distinct compositions are not predicted by population synthesis models as gas accretion should be enhanced at larger orbital separations due to the increased Hill radius and the cooler local gas temperatures that facilitate gas accretion \citep[e.g.][]{Raymond_2018,Burn_2021}. Indeed, most Kepler multi-planet systems are not ``inverted'' \citep[i.e. $0.65\pm 0.4$\% of Kepler multis;][]{Weiss_2018}. Comparative studies of the atmospheric chemistry of both planets in inverted planet pairs will allow for direct tests of sub-Neptune formation models and may help shed light on their potentially unique formation pathway.

\subsection{Implications of the TOI-1266 System for the Emergence of the M Dwarf Radius Valley} \label{sect:radval}
Recall that both TOI-1266 discovery papers reported a radius for TOI-1266 c that was suggestive of a terrestrial composition (i.e. $\sim 1.6\, \mathrm{R}_{\oplus}$; \citetalias{Demory_2020,Stefansson_2020}). If true, the TOI-1266 planets would span the radius valley and serve as an important testbed for radius valley emergence models \citep[e.g.][VanWyngarden et al. in prep.]{Owen_2020}. However, our mass and revised radius measurements indicate that TOI-1266 c is not consistent with a terrestrial composition such that the planet pair does not span the radius valley (c.f. Figure~\ref{fig:mr}). We therefore do not use the multi-transiting TOI-1266 system to test for consistency with atmospheric escape and primordial radius valley formation models, although we do test their plausibility in the following subsections.

\subsubsection{H/He Gas Accretion} \label{sect:accretion}
Here we confirm that the low masses of TOI-1266 b and c are able to accrete enough H/He to explain their masses and radii. The maximum envelope mass that a core of mass $M_{\mathrm{core}}$ can accrete is the maximally cooled isothermal mass $M_{\mathrm{iso}}$ \citep{Lee_2015}. This limit leads to the possibility of a primordial radius valley wherein low mass cores \citep[$\lesssim 1-2\, \mathrm{M}_{\oplus}$;][]{Lee_2022} cannot accrete massive envelopes and consequently form the population of super-Earths. Conversely, more massive cores do accrete H/He envelopes to become sub-Neptunes. This primordial radius valley model has been shown to reproduce the empirical radius valley around FGK stars \citep{Lee_2022}. 

We calculate $M_{\mathrm{iso}}$ for TOI-1266 b and c following the framework and assumptions outlined in \cite{Lee_2022}. Assuming in-situ formation, we find that for gas accretion timescales $\geq 1$ Myr, both planets are capable of accreting $X_{\mathrm{iso}} = M_{\mathrm{iso}}/M_{\mathrm{core}}$ greater than their observed  $X_{\mathrm{env}}$. If we relax the assumption of in situ formation, as is suggested by the system's proximity to a low-order MMR via convergent disk migration \citep[e.g.][]{Cresswell_2006,Tamayo_2017}, then formation in the outer planetary system only serves to increase both planet's maximum envelope masses as accretion is enhanced at lower gas temperatures. We conclude that both planets are capable of accreting massive envelopes that are in excess of what their present-day masses and radii suggest.

Given that H/He accretion onto the TOI-1266 planet cores is plausible, here we estimate their expected initial envelope mass fractions $X_{\mathrm{env,0}} \approx 0.02\, (M_{\mathrm{core}}/\mathrm{M}_{\oplus})^{4/5}\, (T_{\mathrm{eq}}/10^3\, \mathrm{K})^{-1/4}$ \citep{Ginzburg_2016}. We calculate that both planets formed with $X_{\mathrm{env,0}}\sim 6-8$\%, which is plausible given that these values exceed their present-day envelope mass fractions of 5.5\% and 3.5\%, respectively.

\subsubsection{Hydrodynamic Atmospheric Escape Modeling} \label{sect:xuv}
Given that both planets are capable of accreting substantial H/He envelopes, here we test the plausibility that those primordial envelopes can be retained. We consider thermally-driven escape by XUV-driven photoevaporation and core-powered mass loss, both separately and simultaneously. We model these processes using the numerical model \texttt{IsoFATE} (Cherubim \& Wordsworth in prep) wherein the XUV-driven component follows the prescription detailed in \cite{Cherubim_2023}, while the CPML model follows from \cite{Gupta_2020}. Here we provide a brief summary.

\begin{figure}
    \centering
    \includegraphics[width=0.9\hsize]{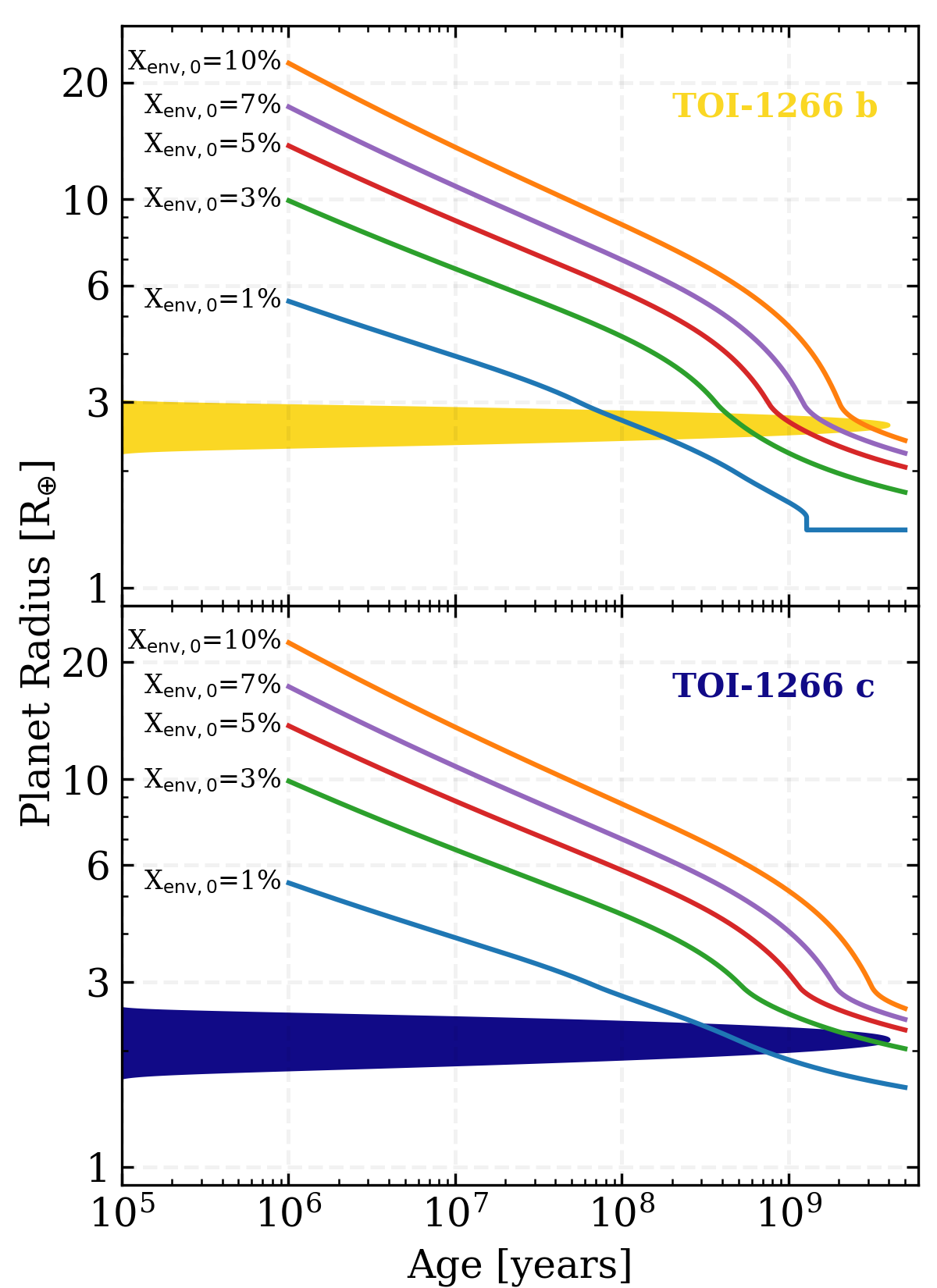}
    \caption{Results from hydrodynamic escape models that combine XUV photoevaporation, core-powered mass loss, and thermal contraction for TOI-1266 b (top) and c (bottom). We present models for different initial envelope mass fractions $X_{\mathrm{env,0}} \in [0.01,0.1]$. The shaded regions in each panel depict the planet's radius posterior. We note that the apparent discontinuity beyond $10^9$ years in our model of TOI-1266 b with $X_{\mathrm{env,0}}=0.01$ occurs when complete atmospheric loss is achieved, leaving behind a bare core.}
    \label{fig:xuv}
\end{figure}

We adopt models of planetary structure following conventional assumptions of a clear H/He envelope at solar metallicity with a deep convective zone beneath an outer, isothermal radiative zone at $T_{\mathrm{eq},i}$. The star remains in the saturated activity regime for 500 Myr with $L_{\mathrm{XUV}}/L_{\mathrm{bol}} = 10^{-3}$ before decaying as a powerlaw, consistent with expectations from semi-empirical MUSCLES spectra \citep{Peacock_2020}. Our model also includes the thermal contraction \citep{Lopez_2014} and Ly-$\alpha$ cooling following the recombination-limited escape regime \citep{MurrayClay_2009}. We do not explicitly model gas accretion onto the planetary cores and instead consider a grid of initial envelope mass fractions $X_{\mathrm{env,0}} \in [0.01,0.1]$.

We evolve the atmospheres under the aforementioned effects and present our results in Figure~\ref{fig:xuv}. We find that neither planet is completely stripped after 5 Gyrs for all but the smallest $X_{\mathrm{env,0}}$ of 0.01. Yet despite retaining some H/He, we find that it is difficult to reconcile the observed radius of planet b with any model that includes XUV-driven photoevaporation, either solely or alongside core-powered mass loss, because its envelope is efficiently lost, leaving behind $X_{\mathrm{env,b}}<5.5$\%. Meanwhile, the observed radius of TOI-1266 c can be recovered when $X_{\mathrm{env,0,c}}\gtrsim 0.03$.

We conclude that TOI-1266 b is more susceptible to atmospheric escape compared to c. While there are instances of our escape model that are consistent with the mass and radius of TOI-1266 c, the same mass loss effect on planet b always results in too much atmospheric loss. Therefore, because a H/He layer is needed to explain the mass and radius of planet b, our atmospheric escape model strongly suggests that planet c is inconsistent with a purely H/He-enveloped terrestrial planet without a substantial volatile mass fraction. Our escape model therefore supports the hypothesis that TOI-1266 c is water-rich.

\subsubsection{Reconciling with Expectations from Hydrodynamic Escape}
Our hydrodynamic models are based on what may be referred to as a `standard picture' of atmospheric escape of H/He envelopes atop terrestrial cores. Such models have successfully reproduced the observed radius valley around FGK stars and ruled out core compositions that are either water or iron-rich \citep{Owen_2017,Wu_2019,Gupta_2020}. Our results indicate that the TOI-1266 planets are inconsistent with this standard picture.

For example, our models suggest that TOI-1266 b is particularly susceptible to XUV-driven escape. This planet may therefore require an alternative explanation that prevented its H/He envelope from being completely stripped during the first $\sim 500$ Myrs. Speculative explanations include the possibility of its late arrival at its current orbit after the star's saturated XUV phase or by the inhibition of hydrodynamic escape by either i) atmospheric shielding by high albedo hazes or ii) the sequestration of H$_2$ in the planet's surface magma that is subsequently outgassed at late times as the core cools \citep{Chachan_2018}. Other amendments to the so-called `standard picture' are the inclusion of additional chemical components that impact atmospheric thermal structure. This includes silicate vapour, which introduces a near-surface radiative layer and alters a planet's size compared to H/He gas alone \citep{Misener_2022}. It may also include a substantial volatile mass fraction if the planet forms water-rich from beyond the snow line \citep[e.g.][]{Venturini_2020}. 

\subsection{Lines of Evidence of a Unique Formation Pathway}
\subsubsection{Low Density Sub-Neptunes Around M Dwarfs}
There is an outstanding tension between the mass-radius relations of sub-Neptune-sized planets based on RV-derived versus TTV-derived masses \citep{Jontof_2014,Steffen_2016}. Specifically beyond orbital periods of 11 days, RV-derived planetary masses appear to be systematically larger than those measured with TTVs, across all radii $<8\, \mathrm{R}_{\oplus}$ \citep{Mills_2017}. This tension is exhibited at the planet population level rather than in individual systems that have high quality RV and TTV data \citep{Malavolta_2017,Borsato_2019}. A plausible explanation for the population level discrepancies is that TTV signal detections are increasingly likely in compact systems with small period ratios (i.e. $\lesssim 2$), particularly those close to low-order MMRs. This introduces a bias in which TTVs preferentially probe compact planet pairs with small period ratios compared to RV measurements. If it is true that sub-Neptune pairs preferentially form with tight spacings due to gas damping \citep{Dawson_2016}, then TOI-1266 b and c may be tracing such a formation pathway given their low period ratio of $P_c/P_b=1.72$. However, the crux of this argument remains that these planets appear to have distinct compositions of gas-enveloped and water-rich, which is challenging to reconcile with this picture of formation.

It is worth highlighting that TOI-1266 is somewhat unique in that it is an M dwarf system that exhibits a low period ratio and has precisely measured RV masses. While the comparison of our RV-derived versus TTV-derived masses in the system is premature due to the lack of a robust TTV detection from TESS, we highlight that these planets are among the lowest density planets transiting M dwarfs between $1.7-3\, \mathrm{R}_{\oplus}$ (Figure~\ref{fig:mr}). This is consistent with multi-planet systems exhibiting lower planet densities \citep{Rodriguez_2023} and suggests that the TOI-1266 planets are more consistent with the aforementioned population of systematically low-density TTV planets that may form via a distinct pathway than the bulk of other small planets around M dwarfs. High-precision ground-based measurements capable of detecting smaller TTVs than TESS can enable a more complete comparison of the RV versus TTV-derived masses in the TOI-1266 system (Greklek-McKeon et al. in prep.).

\subsubsection{Impact of Metallicity on Planet Composition}
Although uncertain, TOI-1266 appears to have a sub-solar metallicity ([Fe/H] $\in [-0.31,-0.08]$ dex), which puts the planets' low bulk densities further at odds with predictions from planet formation models. This is because high metallicity environments are where we expect massive cores to form rapidly enough to trigger the accretion of a substantial gas envelope \citep{Dawson_2015}. With $X_{\mathrm{env},b}=5.5\pm 0.7$\%, TOI-1266 b's envelope mass fraction is one of the highest among planets around M dwarfs and is comparable to the puffy sub-Neptune GJ 1214 b \citep[$X_{\mathrm{env}} = 5.24^{+0.30}_{-0.29}$\%;][]{Cloutier_2021}. However, the large envelope mass of GJ 1214 b is easily reconciled with the host star's high metallicity \citep[$0.29\pm 0.12$;][]{newton14}, which likely facilitated rapid core formation and makes GJ 1214 b a rare gas-enveloped planet around mid-to-late M dwarfs \citep{Ment_2023}. The TOI-1266 planets do not benefit from forming in a similarly metal-rich environment, which adds to the system's complexity. Although the impact of photoevaporative mass loss may be less severe around metal-poor stars, which have been argued to exhibit lower levels of stellar activity \citep{See_2021}.

\subsubsection{Inside-Out Planet Formation} \label{sect:insideout}
The concept of inside-out planet formation offers a plausible explanation for the unique architecture of the TOI-1266 system; including the existence of its ``inverted'' inner planet pair and a third outer planet. Inside-out formation features planetary core formation at the pressure maximum of the inner boundary of the disk's magnetorotational instability dead zone \citep{Chatterjee_2014}. Although the exact location of a dead zone's inner boundary (DZIB) is sensitive to the disk parameters, it is typically located within $\sim 1$ au \citep{Matsumura_2005,Mohanty_2013}, which is comparable to the expected location of the water snow line in dusty protoplanetary disks around low mass stars \citep{Mulders_2015}. Pebble drift in disks around low mass T-Tauri stars proceeds faster and at smaller grain sizes than in more massive disks around typical T-Tauri disks \citep{Pinilla_2013}. This allows massive planetary cores to form rapidly out of the local pebble isolation mass that is enhanced at the DZIB \citep{Bitsch_2018}. In the inside-out formation picture, after its formation at the DZIB, this first planetary core may undergo inward type I migration or, if sufficiently massive, can open a gap in the disk that results in the DZIB retreating to larger orbital separations. In either case, the planetary core and DZIB become spatially decoupled, which then allows for additional pebbles to pile up at the now-cleared DZIB, producing a second planetary core from a depleted pebble reservoir. This process can proceed to form additional planets sequentially.

Many of the features of the TOI-1266 system appear to fit this picture if tuned somewhat finely. While neither planet b nor c are massive enough to induce DZIB retreat, they are both sufficiently massive to undergo type I migration \citep{Burn_2021}. The distinct bulk compositions of TOI-1266 b and c may be naturally explained if the inner planet b formed during the earliest stages of the system's lifetime when the DZIB was located inside of the snow line and the gas surface density was sufficient for substantial H/He accretion. After approaching the local pebble isolation mass, TOI-1266 b could have migrated inward, therefore leaving space for c to form at the DZIB. If planet c's formation timescale is sufficiently long such that the disk gas becomes depleted and the luminosity of the pre-main-sequence TOI-1266 drops, then it is conceivable that while planet c may have formed at the same location as b, the environment in which it formed may have been gas-poor and beyond the snow line \citep{Martin_2012}. The existence of the massive outer planet candidate d, if already formed by an independent process, would have also contributed to depleting the reservoir of pebbles available to form planet c. This could naturally explain why the sandwiched planet c is the lowest mass planet in the system \citep{Pritchard_2023}.

We note that the reliance on the DZIB, as opposed to another location in the disk, may be critical in explaining the architecture of the TOI-1266 system. This is because our measured planet masses (i.e. $\lesssim 4.6\, \mathrm{M}_{\oplus}$) are likely too low to induce pressure maxima that are independent of a DZIB\footnote{Planetary cores typically require masses $\gtrsim 10\, \mathrm{M}_{\oplus}$, depending on assumed disk parameters, to form pressure bumps capable of trapping pebbles \citep{Rosotti_2016}.}. The TOI-1266 planets are also too low-mass to open gaps and induce DZIB retreat. Consequently, type 1 migration would likely be responsible for moving each planet away from the DZIB, which is required to enable sequential planet formation. 

\subsection{Three-body Dynamical Considerations}
\begin{figure}
    \centering
    \includegraphics[width=0.99\hsize]{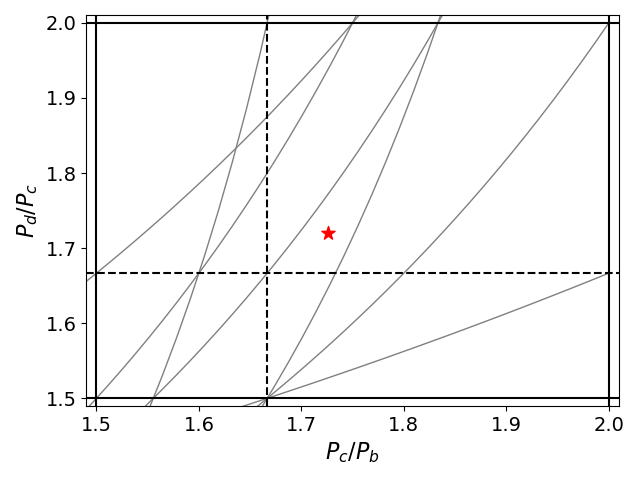}
    \caption{Period ratios of the TOI-1266 planets b, c, and candidate d, in relation to nearby low-order two- and three-body mean motion resonances (MMRs). Locations of 3:2 and 2:1 first-order two-body MMRs are shown by the solid horizontal and vertical lined. Dashed horizontal and vertical lines indicate the location of the second-order 5:3 MMRs. Grey lines indicate the location of strongest nearby three-body MMRs, occurring where $ j' n_d  + (1'-j') n_c  = j n_c +(1-j) n_b$,  $5 n_d  -3 n_c  = j n_c +(1-j) n_b$, or $ j' n_d  + (1'-j') n_c  = 5 n_c -3 n_b$ for $j,j' = 2,3$.}
    \label{fig:resonance}
\end{figure}

In Figure~\ref{fig:resonance}, we show that the orbital configuration of the TOI-1266 does not lie close to any low-order two or three-body MMRs. The nominal period ratio of the inner planet pair, $P_c/P_b$, places it somewhat close to the second-order 5:3 MMR with $\Delta_{bc}= 3(P_c/P_b)/5-1 = 0.035$. However, using the resonance width formula given in Equation 5 of \cite{Hadden_2018}, we determine that, for the RV-measured planets masses and eccentricity constraints from the lack of TTVs in the TESS data, the planet pair is outside of the 5:3 MMR. We reach the same conclusion for the c-d pair, which sits at a similar distance $\Delta_{cd} = 0.032$ from the 5:3 MMR. Figure~\ref{fig:resonance} also shows that the system does not sit at any low-order three-body MMR, which are shown as the loci of points where

\begin{equation}
\frac{j'}{P_d} +\frac{(k'-j')}{P_c} = \frac{(j-k)}{P_b} - \frac{j}{P_c},
\end{equation}

\noindent for the combinations $(j,k,j',k')=$ $\{(3,1,3,1), (2,1,2,1), (3,1,5,2), (2,1,5,2), (5,2,3,1),(5,2,2,1)\}$. The strength of such three-body MMRs scale, to leading order in eccentricity, as $e^{k+k'-2}$, and so these combinations represent the strongest nearby three-body MMRs up to order $e^2$ in the vicinity of the system’s current orbital configuration. We conclude based on the current data that the system is likely not in a resonant chain, but more precise transit timing measurements may be able to directly confirm whether the two- and three-body resonant angles are librating and therefore definitively rule out a resonant configuration.

\section{Summary and Conclusions} \label{sect:summary}
We have presented an RV plus transit modeling analysis of the TOI-1266 planetary system using multiple years of data from TESS and the HARPS-N spectrograph. Our main conclusions are summarized below.

\begin{itemize}
    \item Both planets, but TOI-1266 c in particular, exhibit increased transit depths from the TESS primary mission to the first extended mission. Their revised radii are $R_{p,b}=2.62\pm 0.11\, \mathrm{R}_{\oplus}$ and $R_{p,c}=2.13\pm 0.12\, \mathrm{R}_{\oplus}$. The planet pair does not span the radius valley.
    \item We measure planetary masses of $M_{p,b}=4.23\pm 0.69\, \mathrm{M}_{\oplus}$, $M_{p,c}=2.88\pm 0.80\, \mathrm{M}_{\oplus}$, and find evidence for an outer planet candidate in the system with 
    $M_{p,d}\sin{i}=4.59^{+0.96}_{-0.94}\, \mathrm{M}_{\oplus}$.
    \item Despite both period ratios $P_c/P_b$ and $P_d/P_c$ being close to 5:3, we demonstrate that the system likely does not form a resonant chain.
    \item TOI-1266 b and c are among the lowest density sub-Neptunes known around M dwarfs  ($1.7-3\, \mathrm{R}_{\oplus}$).
    \item TOI-1266 b requires a H/He envelope to explain its mass and radius ($X_{\mathrm{env},b}=5.5\pm 0.7$\%) while TOI-1266 c is consistent with having a water mass fraction whose value is expected from models of formation beyond the snow line (WMF$_c=59\pm 14$\%). The water-rich interpretation of planet c is supported by hydrodynamic escape models, although a H/He envelope cannot be ruled out without atmospheric characterization.
    \item If the aforementioned compositions are true, then TOI-1266 would be the only known M dwarf system whose larger planet has a smaller orbital separation (i.e. an ``inverted'' system) and that features planets with distinct bulk compositions.
    \item The system's unique architecture and distinct planet compositions may be explained by expectations from inside-out sequential planet formation.
\end{itemize}

The unique architecture of the TOI-1266 planetary system presents unique challenges to planet formation models that have been largely successful at explaining properties of the small planet population around M dwarfs. Progress toward validating the inside-out picture of planet formation, or the development of alternative theories, will rely on more stringent constraints on the planets' bulk compositions. Given the degeneracies in inferring accurate bulk compositions from small planets from masses and radii alone, comparative atmospheric planetology will be needed to firmly distinguish between water versus H/He-rich compositions and to shed light on the formation pathway of this system.

\section*{Acknowledgements}
  We thank Jon Jenkins and the SPOC team for their investigation of the TESS light curve upon the discovery of the transit depth discrepancy. In particular, the team's careful reassessment of the effects of crowding on the light curve. We thank Gudmundur Stef\'ansson for sharing the ARCTIC/APO light curve data. We also thank Farzana Meru and Madison VanWyngarden for insight discussions.

  R.C. acknowledges support from the Natural Sciences and Engineering Council of Canada (NSERC) and the Banting Postdoctoral Fellowship Program administered by the Government of Canada.

M.Pi. acknowledges the financial support from the ASI-INAF Addendum no. 2018-24-HH.1-2022 ``Partecipazione italiana al Gaia DPAC – Operazioni e attivit\`a di analisi dati''.

  F.R. is funded by the University of Exeter’s College of Engineering, Maths and Physical Sciences, UK.
  
  This paper includes data collected with the TESS mission, obtained from the MAST data archive at the Space Telescope Science Institute (STScI). Funding for the TESS mission is provided by the NASA Explorer Program. STScI is operated by the Association of Universities for Research in Astronomy, Inc., under NASA contract NAS 5–26555.

  Based on observations made with the Italian Telescopio Nazionale Galileo (TNG)   operated by the Fundaci\'on Galileo Galilei (FGG) of the Instituto Nazionale di Astrofisica (INAF) at the Observatorio del Roque de los Muchachos (La Palma, Canary Islands, Spain).

  The HARPS-N project has been funded by the Prodex Program of the Swiss Space Office (SSO), the Harvard University Origins of Life Initiative (HUOLI), the Scottish Universities Physics Alliance (SUPA), the University of Geneva, the Smithsonian Astrophysical Observatory (SAO), the Italian National Astrophysical Institute (INAF), the University of St Andrews, Queens University Belfast, and the University of Edinburgh.

  This work has made use of data from the European Space Agency (ESA) mission Gaia (\url{https://www.cosmos.esa.int/gaia}), processed by the Gaia Data Processing and Analysis Consortium (DPAC,
  \url{https://www.cosmos.esa.int/web/gaia/dpac/consortium}). Funding for the DPAC has been provided by national institutions, in particular, the institutions participating in the Gaia Multilateral Agreement.

\section*{Data Availability}
This paper includes data collected by the TESS mission, which is publicly available from the Mikulski Archive for Space Telescopes (MAST) at the Space Telescope Science Institute (STScI) \url{https://mast.stsci.edu}. The HARPS-N data underlying this article are available in the article and in its online supplementary material.



\bibliographystyle{mnras}
\bibliography{refs_master} 






\begin{landscape}
\begin{table}
\centering
\caption{HARPS-N Spectroscopic Time Series of TOI-1266. \label{tab:rv}}
\begin{tabular}{rrrrrrrrrrrrrrrrr}
\toprule
Time & LBL RV & $\sigma_{RV}$ & $B$-band RV & $\sigma_{B,RV}$ & $V$-band RV & $\sigma_{V,RV}$ & $R$-band RV & $\sigma_{R,RV}$ & LBL DLW & $\sigma_{DLW}$ & $H\alpha$ & $\sigma_{H\alpha}$ & FWHM & $\sigma_{FWHM}$ & BIS & $\sigma_{BIS}$ \\ 
$[$BJD - 2,457,000$]$ & [m/s] & [m/s] & RV [m/s] & [m/s] & RV [m/s] & [m/s] & RV [m/s] & [m/s] & [m$^2$/s$^2$] & [m$^2$/s$^2$] &&& [km/s] & [km/s] & [km/s] & [km/s] \\
\midrule

 1977.51159 &    -41641.20 &            1.52 &  -41641.87 &         18.29 &  -41639.81 &          6.71 &  -41642.93 &          4.63 &        39055.89 &            1443.00 &   -0.03 &       0.00 &        4.33 &           0.01 &       0.04 &          0.01 \\
 1978.51986 &    -41643.96 &            1.56 &  -41653.21 &         18.29 &  -41639.30 &          6.71 &  -41645.15 &          4.63 &        39670.23 &            1494.20 &   -0.04 &       0.00 &        4.35 &           0.01 &       0.04 &          0.01 \\
 1979.53014 &    -41636.88 &            1.82 &  -41630.25 &         18.29 &  -41636.18 &          6.71 &  -41637.34 &          4.63 &        31908.05 &            1826.00 &   -0.04 &       0.00 &        4.34 &           0.01 &       0.04 &          0.01 \\
 1983.53921 &    -41644.95 &            3.98 &  -41689.41 &         18.29 &  -41626.64 &          6.71 &  -41651.29 &          4.63 &         6525.70 &            4045.92 &   -0.02 &       0.00 &        4.35 &           0.03 &       0.01 &          0.03 \\
 1985.48122 &    -41638.76 &            1.88 &  -41634.55 &         18.29 &  -41633.94 &          6.71 &  -41642.21 &          4.63 &        20536.88 &            1905.23 &   -0.02 &       0.00 &        3.27 &           0.01 &       0.00 &          0.01 \\
\bottomrule    
\multicolumn{17}{l}{For conciseness, only a subset of rows are depicted here to illustrate the table's contents. The entirety of this table is provided in the arXiv source code.} \\
\end{tabular}
\end{table}
\end{landscape}

\clearpage
\newpage

{ \itshape \footnotesize \noindent
$^{1}$Department of Physics \& Astronomy, McMaster University, 1280 Main St West, Hamilton, ON, L8S 4L8, Canada \\
$^{2}$Center for Astrophysics $\vert$ Harvard \& Smithsonian, 60 Garden Street, Cambridge, MA 02138, USA \\
$^{3}$Division of Geological and Planetary Sciences, California Institute of Technology, Pasadena, CA 91125, USA \\
$^4$Earth and Planetary Science, Harvard University, 20 Oxford St, Cambridge, MA 02138, USA \\
$^{5}$Department of Physics and Kavli Institute for Astrophysics and Space Research, Massachusetts Institute of Technology, 77 Massachusetts Ave, Cambridge, MA 02139, USA \\
$^{6}$Canadian Institute for Theoretical Astrophysics, University of Toronto, 60 St. George Str, Toronto, ON, M5S 3H8, Canada \\
$^7$Universit\'e de Montr\'eal, D\'epartement de Physique, iREx, Montr\'eal, QC, H3C 3J7, Canada \\
$^8$Observatoire du Mont-M\'egantic, Universit\'e de Montr\'eal, Montr\'eal, QC, H3C 3J7, Canada \\
$^{9}$School of Physics and Astronomy, University of Birmingham, Edgbaston, Birmingham B15 2TT, UK \\
$^{10}$Astrophysics Group, University of Exeter, Exeter EX4 2QL, UK \\
$^{11}$Instituto de Astrofisica de Canarias, Via Lactea sn, 38200, La Laguna, Tenerife, Spain \\
$^{12}$Fundacion Galileo Galilei - INAF, Rambla J.A.Fernandez P., 7, 38712 (S.C.Tenerife), Spain \\
$^{13}$INAF – Osservatorio Astrofisico di Torino, via Osservatorio 20, 10025 Pino Torinese, Italy \\
$^{14}$University of Li\`ege, Quartier Agora, All\'ee du six Ao\^ut 19C, 4000 Li\`ege, Belgium \\
$^{15}$SUPA, School of Physics and Astronomy, University of St Andrews, North Haugh, St Andrews KY169SS,UK \\ 
$^{16}$DTU Space, National Space Institute, Technical University of Denmark, Elektrovej 328, DK-2800 Kgs. Lyngby, Denmark \\
$^{17}$Israel Institute for Advanced Studies, The Hebrew University of Jerusalem, Edmond J. Safra Campus, Givat Ram, Jerusalem 91904, Israel \\
$^{18}$Observatoire Astronomique de l'Universit\'e de Gen\`eve, 51 chemin des Maillettes, 1290 Versoix, Switzerland \\
$^{19}$INAF - Osservatorio Astronomico di Cagliari, via della Scienza 5, I-09047 Selargius, Italy \\
$^{20}$Dip. di Fisicae Astronomia Galileo Galilei –Universit`a di Padova, Vicolo dell’Osservatorio 2,I-35122 Padova,Italy \\
$^{21}$SUPA, Institute for Astronomy, University of Edinburgh, Blackford Hill, Edinburgh, EH9 3HJ, Scotland, UK \\
$^{22}$Centre for Exoplanet Science, University of Edinburgh, Edinburgh, UK \\
$^{23}$Astrophysics Research Centre, School of Mathematics and Physics, Queen's University Belfast, Belfast, BT7 1NN, UK 
}

\bsp	
\label{lastpage}
\end{document}